\documentclass[11pt]{article}
\usepackage{fullpage,amsmath,amsthm,amssymb,bm,hyperref,cleveref,graphicx,url}
\usepackage[linesnumbered,ruled,vlined]{algorithm2e}
\usepackage{authblk,supertabular}

\usepackage{xcolor}
\usepackage{amsthm}
\usepackage{framed}
 
\graphicspath{{./Fig/}}

\newtheorem{theorem}{Theorem}
\newtheorem{lemma}[theorem]{Lemma}
\newtheorem{corollary}[theorem]{Corollary}
\newtheorem{property}[theorem]{Property}
\newtheorem{proposition}[theorem]{Proposition}
\newtheorem{definition}[theorem]{Definition}

\newenvironment{BoxedTheorem}
   { \colorlet{shadecolor}{black!10}\begin{shaded}\begin{theorem}}
   {\end{theorem}\end{shaded}} 
	
	\newenvironment{BoxedDefinition}
   { \colorlet{shadecolor}{black!3}\begin{shaded}\begin{definition}}
   {\end{definition}\end{shaded}} 
	
	\newenvironment{BoxedProposition}
   { \colorlet{shadecolor}{black!3}\begin{shaded}\begin{proposition}}
   {\end{proposition}\end{shaded}}

		\newenvironment{BoxedCorollary}
   { \colorlet{shadecolor}{black!3}\begin{shaded}\begin{corollary}}
   {\end{corollary}\end{shaded}}

\def\pvec#1#2{\left[\begin{array}{l}#1\\#2\end{array}\right]}
\def\pmat#1#2#3#4{\left[\begin{array}{ll}#1 & #2\\ #3 & #4\end{array}\right]}
\def\CH{\mathrm{CH}}
\def\ie{{{\em i.e.},\ }}
\def\calX{\mathcal{X}}
\def\calA{\mathcal{A}}
\def\calB{\mathcal{B}}

\def\bbR{\mathbb{R}}

\def\JB{\mathrm{JB}}
\def\JS{\mathrm{JS}}
\def\TV{\mathrm{TV}}

\def\dmu{\mathrm{d}\mu}

\def\calP{\mathcal{P}}
\def\std{\mathrm{std}}

\def\bbR{\mathbb{R}}

\def\KL{\mathrm{KL}}
\def\TV{\mathrm{TV}}

\def\calE{\mathcal{E}}

\def\calX{\mathcal{X}}

\def\calC{\mathcal{C}}
\def\calM{\mathcal{M}}
\def\eqdef{{:=}}

\def\tr{{\mathrm{tr}}}
\def\st{{\ :\ }}
\def\inner#1#2{{\langle #1,#2\rangle}}

\def\Chernoff{\mathrm{Chernoff}}
\def\Bhat{\mathrm{Bhat}}

\title{On a generalization of the Jensen-Shannon divergence and the JS-symmetrization of distances relying on abstract means\footnote{This work appeared in the journal~\cite{JS-MDPI-2019} (2019). This report further includes new results and extensions including \protect\S\ref{sec:propMJS}.}}

\author{Frank Nielsen\thanks{E-mail: {\tt Frank.Nielsen@acm.org}. Paper web page: \url{https://FrankNielsen.github.io/M-JS/}}}
\affil{Sony Computer Science Laboratories, Inc.}
\affil{Tokyo, Japan}

\date{}

\begin{document}
\maketitle

\begin{abstract}
The Jensen-Shannon divergence is a renown bounded symmetrization of the unbounded Kullback-Leibler divergence which measures the total Kullback-Leibler divergence to the average mixture distribution.
However the Jensen-Shannon divergence between Gaussian distributions is not available in closed-form.
To bypass this problem, we present a generalization of the Jensen-Shannon (JS) divergence using abstract means
which yields closed-form expressions when the mean is chosen according to the parametric family of distributions.
More generally, we define the JS-symmetrizations of any distance using generalized statistical mixtures derived from abstract means.
In particular, we first show that the geometric mean is well-suited for exponential families, and report 
two closed-form formula for
  (i) the geometric Jensen-Shannon divergence between probability densities of the same exponential family,  and
	(ii) the geometric JS-symmetrization of the reverse Kullback-Leibler divergence.
As a second illustrating example, we show that the harmonic mean is well-suited for the scale Cauchy distributions, and report a closed-form formula for the harmonic Jensen-Shannon divergence between scale Cauchy distributions. 
We also define generalized Jensen-Shannon divergences between matrices (e.g., quantum Jensen-Shannon divergences) and consider clustering with respect to these novel Jensen-Shannon divergences.
\end{abstract}

\noindent {\bf Keywords}: Jensen-Shannon divergence,  Jeffreys divergence, resistor average distance, Bhattacharyya distance, Chernoff information, $f$-divergence, Jensen divergence, Burbea-Rao divergence, Bregman divergence,
  abstract weighted mean, quasi-arithmetic mean, mixture family, statistical $M$-mixture, exponential family, Gaussian family, Cauchy scale family, clustering.

\section{Introduction and motivations}

\subsection{Kullback-Leibler divergence and its symmetrizations}

Let $(\calX,\calA)$ be a measurable space~\cite{PM-2008} where $\calX$ denotes the sample space  and $\calA$ the $\sigma$-algebra of measurable events.
Consider a positive measure $\mu$ (usually the Lebesgue measure $\mu_L$ with Borel $\sigma$-algebra $\calB(\bbR^d)$ or the counting measure  $\mu_c$ with power set 
$\sigma$-algebra $2^\calX$).
Denote by  $\calP$ the set of probability distributions.  

The {\em Kullback-Leibler Divergence}~\cite{CT-2012} (KLD for short) $\KL: \calP\times \calP\rightarrow [0,\infty]$  is the most fundamental distance\footnote{In this paper, we call distance a dissimilarity measure which may not be a metric. Many synonyms have been used in the literature like distortion, deviance, divergence, information, etc.}~\cite{CT-2012} between probability distributions, defined by:
\begin{equation}\label{eq:kldpm}
\KL(P:Q) \eqdef \int p\log \frac{p}{q} \dmu,
\end{equation}
where $p$ and $q$ denote the Radon-Nikodym derivatives  of probability measures $P$ and $Q$ with respect to $\mu$ (with $P,Q\ll\mu$).
The KLD expression between $P$ and $Q$ in Eq.~\ref{eq:kldpm} is independent of the dominating measure $\mu$.
For example, the dominating measure can be chosen as $\mu=\frac{P+Q}{2}$.
Appendix~\ref{sec:im} summarizes the various distances and their notations used in this paper.

The KLD is also called the {\em relative entropy}~\cite{CT-2012} because it can be written as the difference of the cross-entropy $h_\times$ minus the entropy $h$:
\begin{equation}
\KL(p:q) = h_\times(p:q)-h(p),
\end{equation}
where  $h_\times$ denotes the {\em cross-entropy}~\cite{CT-2012}:
\begin{equation}
h_\times(p:q)  \eqdef \int p\log \frac{1}{q} \dmu= -\int p\log {q}\dmu,
\end{equation}
and 
\begin{equation}\label{eq:Sm}
h(p)\eqdef \int p\log \frac{1}{p} \dmu=-\int p\log {p}\dmu=h_\times(p:p),
\end{equation}
denotes the {\em Shannon entropy}~\cite{CT-2012}.
Although the formula of the Shannon entropy in Eq.~\ref{eq:Sm} unifies both the discrete case and the continuous case of probability distributions, the behavior of entropy in the discrete case and in the continuous case is very different:
When $\mu=\mu_c$ (counting measure), Eq.~\ref{eq:Sm} yields the {\em discrete Shannon entropy} which is always positive and upper bounded by $\log |\calX|$.
When $\mu=\mu_L$ (Lebesgue measure), Eq.~\ref{eq:Sm}  defines the Shannon {\em differential entropy} which may be negative and unbounded~\cite{CT-2012} (e.g., the differential entropy of the Gaussian distribution $N(m,\sigma)$ is $\frac{1}{2}\log (2\pi e\sigma^2)$).
See also~\cite{DiscontinuityH-2005} for further important differences between the discrete case and the continuous case.

In general, the KLD is an asymmetric distance (\ie $\KL(p:q)\not=\KL(q:p)$, hence the argument separator notation\footnote{In information theory~\cite{CT-2012}, it is customary to use the double bar notation '$p_X\|p_Y$' instead of the comma ',' notation to avoid confusion with joint random variables $(X,Y)$. } using the delimiter ':') that is {\em unbounded} and may even be infinite.
The {\em reverse KL divergence} or {\em dual KL divergence} is defined by:
\begin{equation}\label{eq:kldpmrev}
\KL^*(P:Q) \eqdef \KL(Q:P) = \int q\log \frac{q}{p} \dmu.
\end{equation}

In general, the {\em reverse distance} or {\em dual distance} for a distance $D$ is written as:
\begin{equation}
D^*(p:q) \eqdef  D(q:p).
\end{equation}

One way to symmetrize the KLD is to consider the {\em Jeffreys\footnote{Sir Harold Jeffreys (1891-1989) was a British statistician.} Divergence}~\cite{Jeffreys-2013} (JD):
\begin{equation}
J(p;q) \eqdef \KL(p:q)+\KL(q:p) = \int (p-q)\log \frac{p}{q} \dmu = J(q;p).
\end{equation}
However this symmetric distance is not upper bounded, and its sensitivity can raise numerical issues in applications.
Here, we used the optional argument separator notation ';' to emphasize that the distance is symmetric\footnote{To match the notational convention of the mutual information if two joint random variables in information theory~\cite{CT-2012}.} but not necessarily a metric distance (i.e., violating the triangle inequality of metric distances).

The symmetrization of the KLD may also be obtained using the {\em harmonic mean} instead of the arithmetic mean,
yielding the {\em resistor average distance}~\cite{resistorKL-2001} $R(p;q)$:
\begin{eqnarray}
\frac{1}{R(p;q)} &=& \frac{1}{2} \left( \frac{1}{\KL(p:q)} +  \frac{1}{\KL(q:p)} \right),\\
\frac{1}{R(p;q)} &=& \frac{\left(\KL(p:q)+\KL(q:p)\right)}{2\KL(p:q)\KL(q:p)},\\
 R(p;q) &=& \frac{2\KL(p:q)\KL(q:p)}{J(p;q)}.
\end{eqnarray}

Another famous symmetrization of the KLD is the {\em Jensen-Shannon Divergence}~\cite{Lin-1991} (JSD)    defined by:
\begin{eqnarray}
\JS(p;q) &\eqdef& \frac{1}{2} \left( \KL\left(p:\frac{p+q}{2}\right) +  \KL\left(q:\frac{p+q}{2}\right) \right),\label{eq:jshi}\\
&=& \frac{1}{2}\int \left(p\log \frac{2p}{p+q} +  q\log \frac{2q}{p+q}\right)\dmu.
\end{eqnarray}
This distance can be interpreted as the {\em total divergence to the average distribution} (see Eq.~\ref{eq:jshi}).
The JSD can be rewritten as a {\em Jensen divergence}
(or Burbea-Rao divergence~\cite{BR-2011}) for the negentropy generator $-h$ (called Shannon information):
\begin{equation}
\JS(p;q) = h\left(\frac{p+q}{2}\right)-\frac{h(p) +h(q) }{2}.\label{eq:jsh}
\end{equation}

An important property of the Jensen-Shannon divergence compared to the Jeffreys divergence is that this Jensen-Shannon distance is {\em always} bounded:
\begin{equation}
0 \leq \JS(p:q) \leq \log 2.
\end{equation}
This follows from the fact that
\begin{equation}
\KL\left(p:\frac{p+q}{2}\right)=\int p\log \frac{2p}{p+q}\dmu \leq \int p\log \frac{2p}{p}\dmu = \log 2.
\end{equation}
Last but not least, the square root of the JSD (i.e., $\sqrt{\JS}$) yields a {\em metric distance} satisfying 
the triangular inequality~\cite{Vajda-MetricDivergence-2009,sqrtJS-2014}.
A variational definition of the Jensen-Shannon divergence and its generalization was studied in~\cite{nielsen2021variational}.

The JSD has found applications in many fields like bioinformatics~\cite{JSDBio-2009} and social sciences~\cite{JSDsocial-2013}, just to name a few.
Recently, the JSD has gained attention in the deep learning community with the {\em Generative Adversarial Networks} (GANs)~\cite{GAN-2014}.

In information geometry~\cite{IG-2016,PRIG-2013,EIG-2020}, the KLD, JD and JSD are {\em invariant divergences} which satisfy the property of information monotonicity.
The class of (separable) distances satisfying the information monotonicity are  exhaustively characterized as Csisz\'ar's $f$-divergences~\cite{Csiszar-1967}.
A {\em $f$-divergence} is defined for a convex generator function $f$ strictly convex at $1$
(with $f(1)=f'(1)=0$) by:
\begin{equation}
I_f(p:q)=\int p f\left(\frac{q}{p}\right)\dmu.
\end{equation}

The Jeffreys divergence and the Jensen-Shannon divergence are $f$-divergence for the following $f$-generators:
\begin{eqnarray}
f_J(u) &\eqdef&  (u-1)\log u,\\
f_\JS(u) &\eqdef& \frac{1}{2}\left(  u\log u -(u+1)\log \frac{1+u}{2} \right).
\end{eqnarray}

\subsection{Statistical distances and parameter divergences}
In information and probability theory, the term ``divergence'' informally means a {\em statistical distance}~\cite{CT-2012}.
However in information geometry~\cite{IG-2016}, a divergence has a stricter meaning of being a smooth {\em parametric} distance (called a contrast function in~\cite{Eguchi-1992}) from which a dual geometric structure can be derived~\cite{DivIG-2010,EIG-2020}.
Consider a parametric family of distributions $\{p_{\theta} \st \theta\in\Theta\}$ (e.g., Gaussian family or Cauchy family), where $\Theta$ denotes the parameter space.
Then a statistical distance $D$ between distributions $p_{\theta}$ and $p_{\theta'}$ amount to an equivalent {\em parameter distance}:
\begin{equation}
P(\theta:\theta')  \eqdef D(p_{\theta}:p_{\theta'}).
\end{equation}
For example, the KLD between two distributions belonging to the same exponential family (e.g., Gaussian family) amount to a reverse Bregman divergence for the cumulant generator $F$ of the exponential family~\cite{Bregman-2005,EIG-2020}:
\begin{equation}
\KL(p_{\theta}:p_{\theta'})= B_F^*(\theta:\theta') = B_F(\theta':\theta).
\end{equation}

A {\em Bregman divergence} $B_F$ is defined for a strictly convex and differentiable generator $F$ as:
\begin{equation}\label{eq:bd}
B_F(\theta:\theta') \eqdef F(\theta)-F(\theta')- \inner{\theta-\theta'}{\nabla F(\theta')},
\end{equation}
where $\inner{\cdot}{\cdot}$ is a inner product (usually the Euclidean dot product for vector parameters).

Similar to the interpretation of the Jensen-Shannon divergence (statistical divergence) as a Jensen divergence for the negentropy generator, the {\em Jensen-Bregman divergence}~\cite{BR-2011} $\JB_F$ (parametric divergence JBD) amounts to a Jensen divergence $J_F$ for a strictly convex generator $F:\Theta\rightarrow \bbR$:

\begin{eqnarray}
\JB_F(\theta:\theta') &\eqdef& \frac{1}{2} \left( B_F\left(\theta:\frac{\theta+\theta'}{2}\right) + 
 B_F\left(\theta':\frac{\theta+\theta'}{2}\right) \right),\\
&=& \frac{F(\theta) +F(\theta') }{2} - F\left(\frac{\theta+\theta'}{2}\right) =: J_F(\theta:\theta'),\label{eq:jbsh}
\end{eqnarray}

Let us introduce the handy notation $(\theta_p\theta_q)_\alpha  \eqdef (1-\alpha)\theta_p+\alpha \theta_q$ to denote the {\em linear interpolation} (LERP) of the parameters (for $\alpha\in [0,1]$).
Then we have more generally that the skew Jensen-Bregman divergence $\JB_F^\alpha(\theta:\theta')$ amounts to a skew Jensen divergence $J_F^\alpha(\theta:\theta')$:
\begin{eqnarray}
\JB_F^\alpha(\theta:\theta') &\eqdef&  (1-\alpha) B_F\left(\theta: (\theta\theta')_\alpha\right) + 
 \alpha B_F\left(\theta': (\theta\theta')_\alpha) \right),\\
&=& (F(\theta)F(\theta'))_{\alpha} - F\left((\theta\theta')_\alpha\right) =: J_F^\alpha(\theta:\theta'),\label{eq:sjbsh}
\end{eqnarray}

\subsection{Jeffreys' $\mathrm{J}$-symmetrization and Jensen-Shannon $\JS$-symmetrization of distances}

For any arbitrary distance $D(p:q)$, we can define its  {\em skew J-symmetrization} for $\alpha\in [0,1]$ by:
\begin{equation}
J_D^\alpha(p:q) \eqdef (1-\alpha) D\left(p:q\right) +  \alpha D\left(q:p\right),
\end{equation}
and its
{\em JS-symmetrization} by:
\begin{eqnarray}
\JS_D^\alpha(p:q) &\eqdef&  (1-\alpha)D\left(p:(1-\alpha)p+\alpha q\right) +  \alpha D\left(q:(1-\alpha)p+\alpha q\right),\\
&=& (1-\alpha)D\left(p:(pq)_\alpha\right) + \alpha D\left(q:(pq)_\alpha\right).
\end{eqnarray}
Usually, $\alpha=\frac{1}{2}$, and for notational brevity, we drop the superscript: $\JS_D(p:q):=\JS_D^{\frac{1}{2}}(p:q)$.
The Jeffreys divergence is twice the $J$-symmetrization of the KLD,
 and the Jensen-Shannon divergence is the $\JS$-symmetrization of the KLD.

The $J$-symmetrization of a $f$-divergence $I_f$ is obtained by taking the generator
\begin{equation}
f^J_\alpha(u) = (1-\alpha)f(u)+\alpha f^\diamond(u),
\end{equation}
where $f^\diamond(u)=u f(\frac{1}{u})$ is the {\em conjugate} generator:
\begin{equation}
I_{f^\diamond}(p:q) = I_f^*(p:q) = I_f(q:p).
\end{equation}

The $\JS$-symmetrization of a $f$-divergence
\begin{equation}
I_f^{\alpha}(p:q) \eqdef (1-\alpha)I_f(p:(pq)_\alpha)+\alpha I_f(q:(pq)_\alpha),
\end{equation}
with $(pq)_\alpha=(1-\alpha)p+\alpha q$ is obtained by taking the generator
\begin{equation}
f^\JS_\alpha(u) \eqdef (1-\alpha)f(\alpha u+1-\alpha)+\alpha	 f\left(\alpha+\frac{1-\alpha}{u}\right). 
\end{equation}

We check that we have:
\begin{equation}
I_f^{\alpha}(p:q) = (1-\alpha)I_f(p:(pq)_\alpha)+\alpha I_f(q:(pq)_\alpha) = I_f^{1-\alpha}(q:p)= I_{f^\JS_\alpha}(p:q).
\end{equation}

A family of symmetric distances unifying the Jeffreys divergence with the Jensen-Shannon divergence was proposed in~\cite{SymJensen-2010}.
Finally, let us mention that once we have symmetrized a distance, we may also  metrize it by choosing (when it exists) the largest exponent $\delta>0$ such that
$D^\delta$ becomes a metric distance~\cite{metricBD1-2008,metricBD2-2008,kafka1991powers,osterreicher2003new,Vajda-MetricDivergence-2009}.

\subsection{Paper outline}
The paper is organized as follows:

\begin{itemize}
\item Section~\ref{sec:mix} reports the special case of mixture families in information geometry~\cite{IG-2016} for which the Jensen-Shannon divergence can be expressed as a Bregman divergence (Theorem~\ref{thm:JSDMix}), 
and highlight the lack of closed-form 
formula when considering generic exponential families. 
This observation precisely motivated this work.

\item Section~\ref{sec:GenJS} introduces the generalized Jensen-Shannon divergences using generalized statistical mixtures derived from abstract weighted means
 (Definition~\ref{eq:JSM} and Definition~\ref{eq:JSMDN}), presents the JS-symmetrization of statistical distances, and report a sufficient condition to get bounded JS-symmetrizations (Property~\ref{prop:MJSDUB}). 

\item In \S\ref{sec:GJS}, we consider the calculation of the geometric JSD between members of the same exponential family (Theorem~\ref{thm:GJSD}) and instantiate the formula for the multivariate Gaussian distributions (Corollary~\ref{cor:GJSDN}).
We show that the Bhattacharrya distance can be interpreted as the negative of the cumulant function of likelihood ratio exponential families in~\S\ref{sec:BhatGJS} and introduce the Chernoff information.
We discuss about applications for $k$-means clustering in \S\ref{sec:kmpp}.
In \S\ref{sec:HJS}, we illustrate the method with another example that calculates in closed-form the  harmonic JSD between scale Cauchy distributions (Theorem~\ref{thm:hjsd}).
\end{itemize}

Finally, we wrap up and conclude this work in Section~\ref{sec:concl}.

\section{Jensen-Shannon divergence in mixture and exponential families}\label{sec:mix}

We are interested to calculate the JSD between densities belonging to parametric families of distributions.
A trivial example is when $p=(p_0,\ldots, p_D)$ and $q=(q_0,\ldots, q_D)$ are categorical distributions (finite discrete distributions sometimes called multinoulli distributions): 
The average distribution $\frac{p+q}{2}$ is again a categorical distribution, and the JSD is expressed plainly as:
\begin{equation}
\JS(p,q) = \frac{1}{2} \sum_{i=0}^D \left(p_i\log \frac{2p_i}{p_i+q_i} +  q_i\log \frac{2q_i}{p_i+q_i}\right).
\end{equation}

Another example is when $p=m_{\theta_p}$ and $q=m_{\theta_q}$ both belong to the same {\em mixture family}~\cite{IG-2016} $\calM$:
\begin{equation}
\calM \eqdef \left\{ m_\theta(x) = \left(1-\sum_{i=1}^D \theta_i p_i(x)\right) p_0(x) + \sum_{i=1}^D \theta_i p_i(x) \st \theta_i>0, \sum_i \theta_i<1\right\},
\end{equation}
for linearly independent component distributions $p_0(x), p_1(x),\ldots, p_D(x)$.
We have~\cite{wmixture-2018}:
\begin{equation}\label{eq:klbfm}
\KL(m_{\theta_p}:m_{\theta_q})=B_F(\theta_p:\theta_q),
\end{equation}
where $B_F$ is a Bregman divergence defined in Eq.~\ref{eq:bd} obtained for the convex negentropy generator~\cite{wmixture-2018} 
$F(\theta)=-h(m_\theta)$.
(The proof that $F(\theta)$ is a strictly convex function is not trivial, see~\cite{MCIG-2018}.)

The mixture families include the family of categorical distributions over a finite alphabet $\calX=\{E_0,\ldots, E_D\}$ (the $D$-dimensional probability simplex) since  those categorical
distributions form a mixture family with $p_i(x)\eqdef \Pr(X=E_i)=\delta_{E_i}(x)$.
Beware that mixture families impose to prescribe the component distributions.
Therefore a density of a mixture family is a special case of statistical mixtures (e.g., Gaussian mixture models) with prescribed component distributions. The special mixtures with prescribed components are also called {\em $w$-mixtures}~\cite{wmixture-2018} because they are convex weighted combinations of components.

The remarkable mathematical identity of Eq.~\ref{eq:klbfm} that does not yield a practical formula since $F(\theta)$ is usually not itself available in closed-form.\footnote{Namely, it is available in closed-form when the fixed component distributions have pairwise disjoint supports with closed-form entropy formula.} Worse, the Bregman generator can be non-analytic~\cite{watanabe2004kullback}. 
Nevertheless, this identity is useful for computing the right-sided Bregman centroid (left KL centroid of mixtures) since this centroid is equivalent to the center of mass, and is always {\em independent} of the Bregman generator~\cite{wmixture-2018}.

The mixture of mixtures is also a mixture: 
\begin{equation}
\frac{m_{\theta_p}+m_{\theta_q}}{2}=m_{\frac{\theta_p+\theta_q}{2}} \in\calM.
\end{equation}
Thus we get a closed-form expression for the JSD between mixtures belonging to $\calM$.

\begin{BoxedTheorem}[Jensen-Shannon divergence between $w$-mixtures]\label{thm:JSDMix}
The Jensen-Shannon divergence between two distributions $p=m_{\theta_p}$ and $q=m_{\theta_q}$ belonging to the same mixture family  $\calM$ is expressed as a Jensen-Bregman divergence for the negentropy generator $F$:
\begin{equation}
\JS(m_{\theta_p},m_{\theta_q}) = \frac{1}{2} \left( B_F\left(\theta_p:\frac{\theta_p+\theta_q}{2}\right) + 
B_F\left(\theta_q:\frac{\theta_p+\theta_q}{2}\right)\right).
\end{equation}
This amounts to calculate the Jensen divergence:
\begin{equation}
\JS(m_{\theta_p},m_{\theta_q}) =  J_F(\theta_1;\theta_2)= (F(\theta_1)F(\theta_2))_{\frac{1}{2}} - F((\theta_1\theta_2)_{\frac{1}{2}}),
\end{equation}
where $(v_1v_2)_\alpha\eqdef (1-\alpha)v_1+\alpha v_2$.
\end{BoxedTheorem}

Now, consider distributions  $p=e_{\theta_p}$ and $q=e_{\theta_q}$ belonging
to the same {\em exponential family}~\cite{IG-2016} $\calE$:

\begin{equation}
\calE\eqdef \left\{e_\theta(x)= \exp\left(\theta^\top x -F(\theta)\right)  \st\theta\in\Theta\right\},
\end{equation}
where 
\begin{equation}
\Theta\eqdef \left\{ \theta\in\bbR^D \st \int \exp(\theta^\top x)\dmu <\infty\right\},
\end{equation}
denotes the natural parameter space.
We have~\cite{IG-2016,EIG-2020}:
\begin{equation}
\KL(e_{\theta_p}:e_{\theta_q})=B_F({\theta_q}:{\theta_p}),
\end{equation}
where $F$ denotes the log-normalizer or cumulant function of the exponential family~\cite{IG-2016} (also called log-partition or log-Laplace function):
\begin{equation}
F(\theta):=\log\left(\int \exp(\theta^\top x)\dmu\right).
\end{equation}

However,  
$\frac{e_{\theta_p}+e_{\theta_q}}{2}$ {\em does not} belong to $\calE$ in general, except for the case of the categorical/multinomial family which is both an exponential family and a mixture family~\cite{IG-2016}.

For example, the mixture of two Gaussian distributions with distinct components is {\em not} a Gaussian distribution.
Thus it is not obvious to get a closed-form expression for the JSD in that case.
This limitation precisely motivated the introduction of generalized Jensen-Shannon divergences defined in the next section.
Notice that in~\cite{fdivchi-2014,powerchifdiv-2019}, it is shown how to express or approximate the $f$-divergences using expansions of power $\chi$ pseudo-distances.
These power $\chi$ pseudo-distances can all be expressed in closed-form when dealing with isotropic Gaussians.
This results holds for the JSD since the JSD is a $f$-divergence~\cite{powerchifdiv-2019}.

\section{Generalized Jensen-Shannon divergences}\label{sec:GenJS}

We first define abstract means $M(x,y)$ of two reals $x$ and $y$, and then define generic statistical $M$-mixtures from which generalized Jensen-Shannon divergences are built thereof.

\subsection{Definitions}

Consider an {\em abstract mean}~\cite{Convexity-2006} $M$.
That is, a continuous bivariate function $M(\cdot,\cdot): I\times I\rightarrow I$ on an interval $I\subset\bbR$ that satisfies the following  
{\em in-betweenness} property:
\begin{equation}
\inf\{x,y\} \leq M(x,y) \leq \sup\{x,y\},\quad \forall x,y\in I.
\end{equation}

Using the unique {\em dyadic expansion} of real numbers, we can always build a corresponding {\em weighted mean} $M_\alpha(p,q)$ (with $\alpha\in [0,1]$) 
following the construction reported in~\cite{Convexity-2006} (page~3)
such that $M_0(p,q)=p$ and $M_1(p,q)=q$.
In the remainder, we consider $I=(0,\infty)$.

Examples of common weighted means are:
\begin{itemize}
\item the {\em arithmetic mean} $A_\alpha(x,y)=(1-\alpha)x+\alpha_y$, 
\item the {\em geometric mean} $G_\alpha(x,y)=x^{1-\alpha}y^\alpha$, and 
\item the {\em harmonic mean} $H_\alpha(x,y)=\frac{1}{(1-\alpha)\frac{1}{x}+\alpha\frac{1}{y}}=\frac{xy}{(1-\alpha)y+\alpha x}$.
\end{itemize}

These means can be unified using the concept of {\em quasi-arithmetic means}~\cite{Convexity-2006}  (also called Kolmogorov-De Finetti-Nagumo means):
\begin{equation}
M^h_\alpha(x,y) \eqdef h^{-1}\left( (1-\alpha)h(x)+\alpha h(y) \right),  
\end{equation}
where $h$ is a strictly monotonous and continuous function.
For example, the geometric mean $G_\alpha(x,y)$ is obtained as $M^h_\alpha(x,y)$ for the generator $h(u)=\log (u)$.
R\'enyi used the concept of quasi-arithmetic means instead of the arithmetic mean to define axiomatically the R\'enyi entropy~\cite{Renyi-1961} of order $\alpha$ in information theory~\cite{CT-2012}.

For any abstract weighted mean $M_\alpha$, we can build a statistical mixture called a {\em $M$-mixture} as follows:

\begin{BoxedDefinition}[Statistical $M$-mixture]
The {\em $M_\alpha$-interpolation} $(pq)_\alpha^M$ (with $\alpha\in [0,1]$) 
of densities $p$ and $q$ with respect to a mean $M$ is a $\alpha$-weighted $M$-mixture defined by:
\begin{equation}
(pq)_\alpha^M(x) \eqdef \frac{M_\alpha(p(x),q(x))}{Z_\alpha^M(p:q)},
\end{equation}
where
\begin{equation}
Z_\alpha^M(p:q) =  \int_{t\in\calX} M_\alpha(p(t),q(t)) \dmu(t) =: \left< M_\alpha(p,q)  \right>.
\end{equation}
is the normalizer function (or scaling factor) ensuring that $(pq)_\alpha^M\in \calP$.
(The bracket notation $\left< f \right>$ denotes the integral of $f$ over $\calX$.) 
\end{BoxedDefinition}

The {\em $A$-mixture}  $(pq)_\alpha^A(x)=(1-\alpha)p(x)+\alpha q(x)$ (`A' standing for the arithmetic mean) represents the usual statistical mixture~\cite{mclachlan-2019} (with $Z_\alpha^A(p:q)=1$).
The {\em $G$-mixture} $(pq)_\alpha^G(x)=\frac{p(x)^{1-\alpha}q(x)^\alpha}{Z_\alpha^G(p:q)}$  of two distributions $p(x)$ and $q(x)$ ('G' standing for the geometric mean $G$) is an exponential family of order $1$, see~\cite{EF-2009}:
\begin{equation}
(pq)_\alpha^G(x)=\exp\left( (1-\alpha)p(x)+\alpha q(x)-\log Z_\alpha^G(p:q) \right).
\end{equation}

The two-component $M$-mixture can be generalized to a {\em $k$-component $M$-mixture} with $\alpha\in \Delta_{k-1}$, the $(k-1)$-dimensional standard simplex:
\begin{equation}
(p_1\ldots p_k)_{\alpha}^M \eqdef \frac{p_1(x)^{\alpha_1}\times \ldots \times p_k(x)^{\alpha_k}}{Z_\alpha(p_1,\ldots,p_k)},
\end{equation}
where $Z_\alpha(p_1,\ldots,p_k) \eqdef \int_\calX   p_1(x)^{\alpha_1}\times \ldots \times p_k(x)^{\alpha_k} \dmu(x)$.

For a given pair of distributions $p$ and $q$, the set $\{M_\alpha(p(x),q(x)) \st \alpha\in [0,1]\}$ describes a path in the space of probability density functions called an Hellinger arc~\cite{Gzyl-1995}. 
This density interpolation scheme was investigated for quasi-arithmetic weighted means  in~\cite{GenBhat-2014,Eguchi-2015,Eguchi-2016}.
In~\cite{JSDiv-2019}, the authors study the Fisher information matrix for the $\alpha$-mixture models (using $\alpha$-power means).

We call $(pq)_\alpha^M$ the {\em $\alpha$-weighted $M$-mixture}, thus extending the notion of $\alpha$-mixtures~\cite{AmariIntegration-2007} obtained for power means $P_\alpha$.
Notice that abstract means have also been used to generalize Bregman divergences using the concept of {\em $(M,N)$-convexity} generalizing the Jensen midpoint inequality~\cite{BDCC-2017}.

Let us state a first generalization of the Jensen-Shannon divergence:

\begin{BoxedDefinition}[$M$-Jensen-Shannon divergence]\label{eq:JSM}
For a mean $M$, the skew $M$-Jensen-Shannon divergence (for $\alpha\in [0,1]$) is defined by
\begin{equation}
{\JS^{M_\alpha}(p:q) \eqdef  (1-\alpha)\,\KL\left(p:(pq)_\alpha^M\right)+\alpha\,\KL\left(q:(pq)_\alpha^M\right)}
\end{equation}
\end{BoxedDefinition}

When $M_\alpha=A_\alpha$, we recover the ordinary Jensen-Shannon divergence since $A_\alpha(p:q)=(pq)_\alpha$ (and $Z_\alpha^A(p:q)=1$).

We can extend the definition to the $\JS$-symmetrization of any distance:

\begin{BoxedDefinition}[$M$-$\JS$ symmetrization]\label{eq:JSMD}
For a mean $M$ and a distance $D$, the skew $M$-$\JS$ symmetrization of $D$ (for $\alpha\in [0,1]$) is defined by
\begin{equation}
{\JS^{M_\alpha}_D(p:q) \eqdef  (1-\alpha)D\left(p:(pq)_\alpha^M\right)+\alpha D\left(q:(pq)_\alpha^M\right)}
\end{equation}
\end{BoxedDefinition}

By notation, we have $\JS^{M_\alpha}(p:q)=\JS^{M_\alpha}_\KL(p:q)$.
That is, the arithmetic JS-symmetrization of the KLD is the JSD.

Let us define the {\em $\alpha$-skew $K$-divergence}~\cite{skew-1999,Lin-1991} $K_\alpha(p:q)$ as
\begin{equation}
K_\alpha\left(p:q\right) \eqdef \KL(p:(1-\alpha)p+\alpha q) = \KL(p:(pq)_\alpha),
\end{equation}
where $(pq)_\alpha(x)\eqdef (1-\alpha)p(x)+\alpha q(x)$.
Then the Jensen-Shannon divergence and the Jeffreys divergence can be rewritten~\cite{SymJensen-2010} as
\begin{eqnarray}
\JS\left(p;q\right) &=& \frac{1}{2}\left(K_{\frac{1}{2}}\left(p:q\right) + K_{\frac{1}{2}}\left(q:p\right) \right),\\
J\left(p;q\right) &=&  K_1(p:q)+K_1(q:p),
\end{eqnarray}
since $\KL(p:q)=K_1(p:q)$.
Then $\JS_\alpha(p:q)=(1-\alpha)K_\alpha(p:q)+\alpha K_{1-\alpha}(q:p)$.
Similarly, we can define the generalized skew $K$-divergence:
\begin{equation}
K^{M_\alpha}_D(p:q) \eqdef  D\left(p:(pq)_\alpha^M\right).
\end{equation}

The success of the JSD compared to the JD in applications is partially due to the fact that the JSD is upper bounded by $\log 2$ (and thus can handle distributions with non-matching supports).
So one question to ask is whether  those generalized JSDs are upper bounded or not?

To report a sufficient condition,  let us first introduce the dominance relationship between means:
We say that a mean $M$ dominates a mean $N$ when $M(x,y)\geq N(x,y)$ for all $x,y\geq 0$, see~\cite{Convexity-2006}.
In that case we write concisely $M\geq N$.
For example, the Arithmetic-Geometric-Harmonic (AGH) inequality states that $A\geq G\geq H$.

Consider the term
\begin{eqnarray}
\KL(p:(pq)_\alpha^M) &=& \int p(x)\log \frac{p(x)Z^M_\alpha(p,q)}{M_\alpha(p(x),q(x))} \dmu(x),\\
&=& \log Z^M_\alpha(p,q) + \int p(x)\log \frac{p(x)}{M_\alpha(p(x),q(x))} \dmu(x).
\end{eqnarray}

When mean $M_\alpha$ dominates the arithmetic mean $A_\alpha$, we have
$$
\int p(x)\log \frac{p(x)}{M_\alpha(p(x),q(x))} \dmu(x) \leq \int p(x)\log \frac{p(x)}{A_\alpha(p(x),q(x))} \dmu(x),
$$
and
$$
\int p(x)\log \frac{p(x)}{A_\alpha(p(x),q(x))} \dmu(x) \leq  \int p(x)\log \frac{p(x)}{(1-\alpha)p(x)} \dmu(x)  = -\log (1-\alpha).
$$

Notice that $Z^A_\alpha(p:q)=1$ (when $M=A$ is the arithmetic mean), and we recover the fact that the $\alpha$-skew Jensen-Shannon divergence is upper bounded by $-\log (1-\alpha)$
 (e.g., $\log 2$ when $\alpha=\frac{1}{2}$).

We summarize the result in the following property:

\begin{property}[Upper bound on $M$-JSD]\label{prop:MJSDUB}
The $M$-JSD is upper bounded by $\log \frac{Z^M_\alpha(p,q)}{1-\alpha}$ when $M\geq A$.
\end{property}

Notice that since $\min\{a,b\} \leq M_\alpha(a,b)\leq \max\{a,b\}$, we have
$$
\int \min\{p(x),q(x)\} \dmu \leq Z^M_\alpha(p,q)\leq \int \max\{p(x),q(x)\} \dmu.
$$
But $\min\{a,b\}=\frac{a+b}{2}-\frac{1}{2}|b-a|$ and $\max\{a,b\}=\frac{a+b}{2}+\frac{1}{2}|b-a|$.
Thus we have 
$$
1-\TV(p,q) \leq Z^M_\alpha(p,q)\leq 1+\TV(p,q),
$$
where $\TV(p,q)=\frac{1}{2}\int |q(x)-p(x)| \dmu(x)$ is the total variation distance.
Since $\TV(p,q)\in [0,1]$, we deduce that for any abstract mean, we have $Z^M_\alpha(p,q)\in [0,2]$.

Let us observe that dominance of means can be used to define distances:
For example, the celebrated $\alpha$-divergences
\begin{equation}
I_\alpha({p}:{q}) =  \int \left(\alpha p(x) +(1-\alpha)q(x)-p(x)^\alpha q(x)^{1-\alpha} \right) \dmu(x),\quad \alpha\not\in\{0,1\}
\end{equation}
can be interpreted as a difference of two means, the arithmetic mean and the geometry mean:
\begin{equation}
I_\alpha({p}:{q}) =  \int \left( A_\alpha(q(x):p(x))-G_\alpha(q(x):p(x)) \right) \dmu(x).
\end{equation}
See~\cite{GenAlphaDiv-2020} for more details.

We can also define the generalized Jeffreys divergence as follows:

\begin{BoxedDefinition}[$N$-Jeffreys divergence]\label{eq:JM}
For a mean $N$, the skew $N$-Jeffreys divergence (for $\beta\in [0,1]$) is defined by
\begin{eqnarray}
J^{N_\beta}(p:q) &\eqdef&  N_\beta(\KL\left(p:q\right),\KL\left(q:p\right)).
\end{eqnarray}
\end{BoxedDefinition}

This definition includes the (scaled) {\em resistor average distance}~\cite{resistorKL-2001} $R(p;q)$, obtained for the {\em harmonic mean} $N=H$ for the KLD with skew parameter $\beta=\frac{1}{2}$:
\begin{eqnarray}
\frac{1}{R(p;q)} &=& \frac{1}{2} \left( \frac{1}{\KL(p:q)} +  \frac{1}{\KL(q:p)} \right),\\
R(p;q)&=& \frac{2\KL(p:q)\KL(q:p)}{J(p;q)}.
\end{eqnarray}
In~\cite{resistorKL-2001}, the factor $\frac{1}{2}$ is omitted to keep the spirit of the original Jeffreys divergence.
Notice that
\begin{equation}
R(p;q)=\frac{G(\KL(p:q),\KL(q:p))}{A(\KL(p:q),\KL(q:p))}\leq 1
\end{equation}
since $A\geq G$ (i.e., the arithmetic mean $A$ dominates the geometric mean $G$).

Thus for any arbitrary divergence $D$, we can define the following generalization of the Jensen-Shannon divergence:

\begin{BoxedDefinition}[Skew $(M,N)$-JS $D$ divergence]\label{eq:JSMDN}
The skew $(M,N)$-divergence with respect to weighted means $M_\alpha$ and $N_\beta$ as follows:
\begin{equation}
{\JS}^{M_\alpha,N_\beta}_D(p:q) \eqdef  N_\beta\left(D\left(p:(pq)_\alpha^M\right),D\left(q:(pq)_\alpha^M\right)\right).
\end{equation}
\end{BoxedDefinition}

We now show how to choose the abstract mean according to the parametric family of distributions in order to obtain some closed-form formula for some statistical distances.


\section{Some closed-form formula for the $M$-Jensen-Shannon divergences\label{sec:cfjs}}

Our motivation to introduce these novel families of $M$-Jensen-Shannon divergences 
is to obtain closed-form formula when probability densities belong to some given parametric families $\calP_{\Theta}$.
We shall illustrate the principle of the method to choose the right abstract mean for the considered parametric family, and report corresponding formula for the following two case studies: 
\begin{enumerate}
\item The {\em geometric $G$-Jensen-Shannon divergence} for the {\em exponential families} (\S\ref{sec:GJS}), and 
\item the {\em harmonic $H$-Jensen-Shannon divergence} for the family of {\em Cauchy scale distributions} (\S\ref{sec:HJS}).
\end{enumerate}

Recall that the arithmetic $A$-Jensen-Shannon divergence is well-suited for mixture families (Theorem~\ref{thm:JSDMix}).

\subsection{The geometric Jensen-Shannon divergence: $G$-JSD\label{sec:GJS}}

Consider an exponential family~\cite{EF-2009} $\calE_F$ with log-normalizer $F$:
\begin{equation}
\calE_F=\left\{ p_\theta(x)\dmu=\exp(\theta^\top x-F(\theta))\dmu  \st \theta\in\Theta \right\},
\end{equation}
and natural parameter space 
\begin{equation}
\Theta = \left\{ \theta \st \int_{\calX} \exp(\theta^\top x)\dmu<\infty \right\}.
\end{equation}
The log-normalizer (a log-Laplace function also called log-partition or cumulant function) is a real analytic convex function.

Choose for the abstract mean $M_\alpha(x,y)$  the {\em weighted geometric mean} $G_\alpha$: 
$M_\alpha(x,y)=G_\alpha(x,y)=x^{1-\alpha}y^\alpha$, for $x,y>0$.

It is well-known that the normalized weighted product of distributions belonging to the same exponential family also belongs to this exponential family~\cite{GMM-2019}:
\begin{eqnarray}
\forall x\in\calX,\quad (p_{\theta_1}p_{\theta_2})_\alpha^G(x) &\eqdef &  
\frac{G_\alpha(p_{\theta_1}(x),p_{\theta_2}(x))}{\int G_\alpha(p_{\theta_1}(t),p_{\theta_2}(t)) \dmu(t)}  = \frac{p_{\theta_1}^{1-\alpha}(x) p_{\theta_2}^\alpha(x)}{Z^G_\alpha(p:q)},\\
 &=&  {p_{(\theta_1\theta_2)_\alpha}(x)},
\end{eqnarray}
where the normalization factor 
is 
\begin{equation}
Z^G_\alpha(p:q)=\exp(-J_F^\alpha(\theta_1:\theta_2)),
\end{equation}
for the skew Jensen divergence $J_F^\alpha$ defined by:
\begin{equation}
J_F^\alpha(\theta_1:\theta_2) \eqdef (F(\theta_1)F(\theta_2))_\alpha-F((\theta_1\theta_2)_\alpha).
\end{equation}

Notice that since the natural parameter space $\Theta$ is convex, the distribution $p_{(\theta_1\theta_2)_\alpha}\in\calE_F$ (since $(\theta_1\theta_2)_\alpha\in\Theta$).

Thus  it follows that we have:  
\begin{eqnarray}
\KL\left(p_{\theta}: (p_{\theta_1}p_{\theta_2})_\alpha^G\right) &=& \KL\left(p_{\theta}: p_{(\theta_1\theta_2)_\alpha}\right),\\
&=& B_F((\theta_1\theta_2)_\alpha:\theta).
\end{eqnarray}

%

This allows us to conclude that the {\em $G$-Jensen-Shannon divergence} admits the following closed-form expression between densities belonging to the same exponential family:

\begin{eqnarray}
\JS^G_\alpha(p_{\theta_1}:p_{\theta_2}) &\eqdef& 
(1-\alpha)\, \KL(p_{\theta_1}:(p_{\theta_1}p_{\theta_2})_\alpha^G) + \alpha\, \KL(p_{\theta_2}:(p_{\theta_1}p_{\theta_2})_\alpha^G),\\
&=& (1-\alpha)\, B_F((\theta_1\theta_2)_\alpha:\theta_1) +\alpha\, B_F((\theta_1\theta_2)_\alpha:\theta_2).
\end{eqnarray}

Note that since $(\theta_1\theta_2)_\alpha-\theta_1=\alpha(\theta_2-\theta_1)$ and 
$(\theta_1\theta_2)_\alpha-\theta_2=(1-\alpha)(\theta_1-\theta_2)$, it follows that
$(1-\alpha)B_F(\theta_1: (\theta_1\theta_2)_\alpha)+\alpha B_F(\theta_2: (\theta_1\theta_2)_\alpha)=J_F^\alpha(\theta_1:\theta_2)$.

The {\em dual divergence}~\cite{RefDuality-2015} $D^*$ (with respect to the reference argument)  or {\em reverse divergence} of a divergence $D$ is defined by swapping the calling arguments:
$D^*(\theta:\theta')\eqdef D(\theta':\theta)$.

Thus if we defined the Jensen-Shannon divergence for the dual KL divergence $\KL^*(p:q)\eqdef \KL(q:p)$ 
\begin{eqnarray}
\JS_{\KL^*}(p:q) &\eqdef& \frac{1}{2} \left( \KL^*\left(p:\frac{p+q}{2}\right) +  \KL^*\left(q:\frac{p+q}{2}\right) \right),\\
 &=& \frac{1}{2} \left( \KL\left(\frac{p+q}{2}:p\right) +  \KL\left(\frac{p+q}{2}:q\right) \right),
\end{eqnarray}
then we obtain:

\begin{eqnarray}
{\JS}_{\KL^*}^{G_\alpha}(p_{\theta_1}:p_{\theta_2}) &\eqdef& 
(1-\alpha) \KL((p_{\theta_1}p_{\theta_2})_\alpha^G:p_{\theta_1}) + \alpha \KL((p_{\theta_1}p_{\theta_2})_\alpha^G:p_{\theta_2}),\\
&=& (1-\alpha) B_F(\theta_1:(\theta_1\theta_2)_\alpha) +\alpha B_F(\theta_2:(\theta_1\theta_2)_\alpha)=\JB_F^\alpha(\theta_1:\theta_2),\\
&=& (1-\alpha)  F(\theta_1) + \alpha F(\theta_2) - F((\theta_1\theta_2)_\alpha),\\
&=& J_F^\alpha(\theta_1:\theta_2).
\end{eqnarray}

Note that $\JS_{D^*}\not = {\JS_D}^*$.

In general, the JS-symmetrization for the reverse KL divergence is
\begin{eqnarray}
{\JS}_{\KL^*}(p;q) &=&   \frac{1}{2}\left(  \KL\left(\frac{p+q}{2}:p\right) + \KL\left(\frac{p+q}{2}:q\right) \right),\\
&=&  \int m \log \frac{m}{\sqrt{pq}}\dmu= \int A(p,q) \log \frac{A(p,q)}{G(p,q)}\dmu,
\end{eqnarray}
where $m=\frac{p+q}{2}=A(p,q)$ and $G(p,q)=\sqrt{pq}$. Since $A\geq G$ (arithmetic-geometric inequality), it follows that ${\JS}_{\KL^*}(p;q)\geq 0$.

\begin{BoxedTheorem}[$G$-JSD and its dual $\JS$-symmetrization in exponential families]\label{thm:GJSD}
The $\alpha$-skew $G$-Jensen-Shannon divergence $\JS^{G_\alpha}$ between two distributions $p_{\theta_1}$ and $p_{\theta_2}$ of the same exponential family $\calE_F$ is expressed in closed-form for $\alpha\in (0,1)$ as:
\begin{eqnarray}
\JS^{G_\alpha}(p_{\theta_1}:p_{\theta_2}) &=& (1-\alpha)B_F\left((\theta_1\theta_2)_\alpha:\theta_1\right)
+ \alpha B_F\left((\theta_1\theta_2)_\alpha:\theta_2\right),\\
{\JS}_{\KL^*}^{G_\alpha}(p_{\theta_1}:p_{\theta_2}) &=&   \JB_F^\alpha(\theta_1:\theta_2) =  J_F^\alpha(\theta_1:\theta_2).
\end{eqnarray}
\end{BoxedTheorem}


\subsubsection{Case study: The multivariate Gaussian family}

Consider the {\em exponential family}~\cite{IG-2016,EF-2009} of multivariate Gaussian distributions~\cite{Yoshizawa-1999,SM-2011,EIG-2018} 
\begin{equation}
\left\{N(\mu,\Sigma) \st \mu\in\bbR^d, \Sigma\succ 0\right\}.
\end{equation}
The multivariate Gaussian family is also called the {\em MultiVariate Normal} family in the literature, or MVN family for short.

Let $\lambda \eqdef(\lambda_v,\lambda_M)=(\mu,\Sigma)$ denote the {\em composite} (vector, matrix) parameter of a MVN.
The $d$-dimensional MVN density is given by
\begin{eqnarray}\label{eq:mvnl}
p_\lambda(x;\lambda) &\eqdef&  \frac{1}{(2\pi)^{\frac{d}{2}}\sqrt{|\lambda_M|}}  \exp\left(-\frac{1}{2} (x-\lambda_v)^\top \lambda_M^{-1} (x-\lambda_v)\right),
\end{eqnarray} 
where $|\cdot|$ denotes the matrix determinant.
The natural parameters $\theta$ are also expressed using both a {vector parameter}  $\theta_v$ and a  {matrix parameter}  $\theta_M$ in a compound object $\theta=(\theta_v,\theta_M)$.
By defining the following {\em compound inner product} on a composite (vector,matrix) object
\begin{equation}
\inner{\theta}{\theta'}\eqdef \theta_v^\top \theta_v'+ \mathrm{tr}\left({\theta_M'}^\top\theta_M\right),
\end{equation}
where $\tr(\cdot)$ denotes the matrix trace, we rewrite the MVN density of Eq.\ref{eq:mvnl} in the canonical form of an exponential family~\cite{EF-2009}:
\begin{eqnarray}
p_\theta(x;\theta) &\eqdef& \exp\left(\inner{t(x)}{\theta}-F_\theta(\theta)\right) = p_\lambda(x;\lambda(\theta)),
\end{eqnarray} 
where 
\begin{equation}
\theta=(\theta_v,\theta_M)=\left(\Sigma^{-1}\mu,\frac{1}{2}\Sigma^{-1}\right)=\theta(\lambda)=\left(\lambda_M^{-1}\lambda_v,\frac{1}{2}\lambda_M^{-1}\right),
\end{equation}
 is the {\em compound natural parameter} and 
 \begin{equation}
t(x)=(x,-xx^\top)
\end{equation}
 is the {\em compound sufficient statistic}.
The  function $F_\theta$ is the strictly convex and continuously differentiable log-normalizer defined by:
\begin{equation}
F_\theta(\theta) = \frac{1}{2}\left( d\log\pi -\log |\theta_M|+\frac{1}{2} \theta_v^\top \theta_M^{-1} \theta_v \right),
\end{equation}
The log-normalizer can be expressed using the ordinary parameters, $\lambda=(\mu,\Sigma)$,  as:
\begin{eqnarray}
F_\lambda(\lambda) &=& \frac{1}{2}\left(\lambda_v^\top \lambda_M^{-1}\lambda_v+\log |\lambda_M| + d\log2\pi \right),\\
&=& \frac{1}{2}\left(\mu^\top \Sigma^{-1}\mu +\log |\Sigma| + d\log2\pi \right).
\end{eqnarray}
The {\em moment/expectation parameters}~\cite{IG-2016,EIG-2018} are 
\begin{equation}
\eta=(\eta_v,\eta_M)=E[t(x)]=\nabla F(\theta).
\end{equation}

We report the conversion formula between the three types of coordinate systems (namely, the ordinary parameter $\lambda$, the natural parameter $\theta$ and the moment parameter $\eta$) as follows:
\begin{eqnarray}
\left\{
\begin{array}{ll}
\theta_v(\lambda)=\lambda_M^{-1}\lambda_v=\Sigma^{-1}\mu\\
\theta_M(\lambda)=\frac{1}{2}\lambda_M^{-1}=\frac{1}{2}\Sigma^{-1}
\end{array}
\right.  &\Leftrightarrow &
\left\{
\begin{array}{ll}
\lambda_v(\theta) = \frac{1}{2}\theta_M^{-1}\theta_v=\mu\\
\lambda_M(\theta) = \frac{1}{2}\theta_M^{-1}=\Sigma
\end{array}
\right.\\
\left\{
\begin{array}{ll}
\eta_v(\theta)= \frac{1}{2}\theta_M^{-1}\theta_v\\
\eta_M(\theta)= -\frac{1}{2}\theta_M^{-1}-\frac{1}{4}(\theta_M^{-1}\theta_v)(\theta_M^{-1}\theta_v)^\top
\end{array}
\right.
  &\Leftrightarrow &
	\left\{
\begin{array}{ll}
\theta_v(\eta)=  -(\eta_M+\eta_v\eta_v^\top)^{-1}\eta_v\\
\theta_M(\eta)=  -\frac{1}{2}(\eta_M+\eta_v\eta_v^\top)^{-1}
\end{array}
\right.\\
\left\{
\begin{array}{ll}
\lambda_v(\eta)=  \eta_v = \mu\\
\lambda_M(\eta)=  -\eta_M-\eta_v\eta_v^\top=\Sigma
\end{array}
\right.
  &\Leftrightarrow &
	\left\{
\begin{array}{ll}
\eta_v(\lambda)=   \lambda_v=\mu\\
\eta_M(\lambda)=   -\lambda_M-\lambda_v\lambda_v^\top =-\Sigma -\mu\mu^\top
\end{array}
\right.
\end{eqnarray}

The dual Legendre convex conjugate~\cite{IG-2016,EIG-2018} is 
\begin{equation}
F^*_\eta(\eta)=-\frac{1}{2}\left( \log(1+\eta_v^\top \eta_M^{-1} \eta_v) + \log |-\eta_M| +d (1+\log 2\pi) \right),
\end{equation}
and $\theta=\nabla_\eta F_\eta^*(\eta)$.
We check the Fenchel-Young equality when $\eta=\nabla F(\theta)$ and $\theta=\nabla F^*(\eta)$:
\begin{equation}
F_\theta(\theta) + F^*_\eta(\eta)  -\inner{\theta}{\eta} =0.
\end{equation}

Notice that the log-normalizer can be expressed using the expectation parameters as well:
\begin{equation}
F_\eta(\eta) =  \frac{1}{2} \eta_v^\top(\eta_M-\eta_v\eta_v^\top)^{-1}\eta_v+\frac{1}{2}\log |2\pi (\eta_M-\eta_v\eta_v^\top)|.
\end{equation}

We have $F_\theta(\theta)=F_\lambda(\lambda(\theta))=F_\eta(\eta(\theta))$.

The Kullback-Leibler divergence between two $d$-dimensional Gaussians distributions $p_{(\mu_1,\Sigma_1)}$ and $p_{(\mu_2,\Sigma_2)}$ 
(with $\Delta_\mu=\mu_2-\mu_1$) is
\begin{eqnarray}
\KL(p_{(\mu_1,\Sigma_1)}:p_{(\mu_2,\Sigma_2)}) 
&=& \frac{1}{2} \left\{ \tr(\Sigma_2^{-1}\Sigma_1) + \Delta_\mu^\top \Sigma_2^{-1}\Delta_\mu +  \log \frac{|\Sigma_2|}{|\Sigma_1|} -d \right\} =\KL(p_{\lambda_1}:p_{\lambda_2}).
\end{eqnarray}
We check that $\KL(p_{(\mu,\Sigma)}:p_{(\mu,\Sigma)})=0$ since $\Delta_\mu=0$ and $\tr(\Sigma^{-1}\Sigma)=\tr(I)=d$.
Notice that when $\Sigma_1=\Sigma_2=\Sigma$, we have
\begin{equation}
\KL(p_{(\mu_1,\Sigma)}:p_{(\mu_2,\Sigma)})  = \frac{1}{2}  \Delta_\mu^\top \Sigma^{-1}\Delta_\mu= \frac{1}{2}D_{\Sigma^{-1}}^2(\mu_1,\mu_2),
\end{equation}
that is half the squared Mahalanobis distance for the precision matrix $\Sigma^{-1}$ (a positive-definite matrix: $\Sigma^{-1}\succ 0$), where the Mahalanobis distance is defined for any positive matrix $Q\succ 0$ as follows:
\begin{equation}
D_Q(p_1:p_2) = \sqrt{(p_1-p_2)^\top Q (p_1-p_2)}.
\end{equation}

The Kullback-Leibler divergence between two probability densities of the same exponential families amount to a Bregman divergence~\cite{IG-2016}:
 \begin{equation}
\KL(p_{(\mu_1,\Sigma_1)}:p_{(\mu_2,\Sigma_2)})  = \KL(p_{\lambda_1}:p_{\lambda_2}) = B_F(\theta_2:\theta_1)=B_{F^*}(\eta_1:\eta_2),
\end{equation}
where the Bregman divergence is defined by
\begin{equation}\label{eq:bdip}
B_F(\theta:\theta') \eqdef F(\theta)-F(\theta')-\inner{\theta-\theta'}{\nabla F(\theta')},
\end{equation}
with $\eta'=\nabla F(\theta')$.
Define the canonical divergence~\cite{IG-2016}
\begin{equation}
A_F(\theta_1:\eta_2)=F(\theta_1)+F^*(\eta_2)-\inner{\theta_1}{\eta_2}=A_{F^*}(\eta_2:\theta_1),
\end{equation}
since ${F^*}^*=F$. We have $B_F(\theta_1:\theta_2)=A_F(\theta_1:\eta_2)$.

Now, observe that $p_\theta(0,\theta)=\exp(-F(\theta))$ when $\inner{t(0)}{\theta}=0$.
In particular, this holds for the multivariate normal family.
Thus we have the following proposition

\begin{proposition}
For the MVN family, we have 
\begin{equation}
p_\theta(x;(\theta_1\theta_2)_\alpha)  = \frac{p_\theta(x,\theta_1)^{1-\alpha}p_\theta(x,\theta_2)^{\alpha}}{Z_\alpha^G(p_{\theta_1}:p_{\theta_2})},
\end{equation}
with the scaling normalization factor:
\begin{equation}
Z_\alpha^G(p_{\theta_1}:p_{\theta_2}) = \exp(-J_F^\alpha(\theta_1:\theta_2)) = 
\frac{p_\theta(0;\theta_1)^{1-\alpha} p_\theta(0;\theta_2)^{\alpha} }{p_\theta(0;(\theta_1\theta_2)_\alpha)}.
\end{equation}
\end{proposition}


More generally, we have for a $k$-dimensional weight vector $\alpha$ belonging to the $(k-1)$-dimensional standard simplex:
\begin{equation}
Z_\alpha^G(p_{\theta_1},\ldots p_{\theta_k}) = \frac{\prod_{i=1}^k p_\theta(0,\theta_i)^{\alpha_i} }{p_\theta(0;\bar\theta)},
\end{equation}
where $\bar\theta=\sum_{i=1}^k \alpha_i\theta_i$.

Finally, we state the formulas for the G-JS divergence between MVNs for the KL and reverse KL, respectively:

\begin{BoxedCorollary}[$G$-JSD between Gaussians]\label{cor:GJSDN}
The skew $G$-Jensen-Shannon divergence $\JS_\alpha^G$ and the dual skew $G$-Jensen-Shannon divergence ${\JS^*}_\alpha^G$ between two multivariate Gaussians $N(\mu_1,\Sigma_1)$ and $N(\mu_2,\Sigma_2)$ is
\begin{eqnarray}
{\JS}^{G_\alpha}(p_{(\mu_1,\Sigma_1)}:p_{(\mu_2,\Sigma_2)}) 
&=& (1-\alpha)\KL(p_{(\mu_1,\Sigma_1)}:p_{(\mu_\alpha,\Sigma_\alpha)})+\alpha \KL(p_{(\mu_2,\Sigma_2)}:p_{(\mu_\alpha,\Sigma_\alpha)}),\\
&=& (1-\alpha)B_F((\theta_1\theta_2)_\alpha:\theta_1)+
\alpha B_F((\theta_1\theta_2)_\alpha:\theta_2),\\
{\JS}^{G_\alpha}_*(p_{(\mu_1,\Sigma_1)}:p_{(\mu_2,\Sigma_2)}) &=& (1-\alpha)\KL(p_{(\mu_\alpha,\Sigma_\alpha)}:p_{(\mu_1,\Sigma_1)})
+\alpha \KL(p_{(\mu_\alpha,\Sigma_\alpha)}:p_{(\mu_2,\Sigma_2)}),\\
&=& (1-\alpha)B_F(\theta_1:(\theta_1\theta_2)_\alpha)+
\alpha B_F(\theta_2:(\theta_1\theta_2)_\alpha),\\
&=& J_F(\theta_1:\theta_2),\\
&=& \frac{1}{2}\left((1-\alpha)\mu_1^\top\Sigma_1^{-1}\mu_1+\alpha\mu_2^\top\Sigma_2^{-1}\mu_2-\mu_\alpha^\top\Sigma_\alpha^{-1}\mu_\alpha
+\log \frac{|\Sigma_1|^{1-\alpha}|\Sigma_2|^\alpha}{|\Sigma_\alpha|}\right),\nonumber
\end{eqnarray}
where 

\begin{equation}
\Sigma_\alpha=(\Sigma_1\Sigma_2)_\alpha^\Sigma= \left((1-\alpha)\Sigma_1^{-1}+\alpha \Sigma_2^{-1}\right)^{-1},
\end{equation}
 (matrix harmonic barycenter)
 and 
\begin{equation}
\mu_\alpha=(\mu_1\mu_2)_\alpha^\mu =\Sigma_\alpha \left((1-\alpha)\Sigma_1^{-1}\mu_1+\alpha \Sigma_2^{-1}\mu_2\right).
\end{equation}
\end{BoxedCorollary}

%

 Notice that the {\em $\alpha$-skew Bhattacharyya distance}~\cite{BR-2011}:
\begin{equation}\label{eq:bhat}
B_\alpha(p:q) = -\log \int_\calX p^{1-\alpha} q^\alpha \dmu 
\end{equation}
between two members of the same exponential family amounts to a $\alpha$-skew Jensen divergence between the corresponding natural parameters:
\begin{equation} 
B_\alpha(p_{\theta_1}:p_{\theta_2}) = J_F^\alpha(\theta_1:\theta_2).
\end{equation}

A simple proof follows from the fact that 
\begin{equation}
\int p_{(\theta_1\theta_2)_\alpha}(x)\dmu(x) = 1 =\int \frac{p_{\theta_1}^{1-\alpha}(x)p_{\theta_2}^\alpha(x)}{Z_\alpha^G(p_{\theta_1}:p_{\theta_2})}\dmu(x). 
\end{equation}

Therefore we have
\begin{equation}
\log 1 =  0 = \log \int p_{\theta_1}^{1-\alpha}(x)p_{\theta_2}^\alpha(x)\dmu(x) -\log Z_\alpha^G(p_{\theta_1}:p_{\theta_2}),
\end{equation}
with $Z_\alpha^G(p_{\theta_1}:p_{\theta_2})=\exp(-J_F(p_{\theta_1}:p_{\theta_2}))$.
Thus it follows that
\begin{eqnarray}
B_\alpha(p_{\theta_1}:p_{\theta_2}) &=& -\log \int p_{\theta_1}^{1-\alpha}(x)p_{\theta_2}^\alpha(x)\dmu(x),\\
 &=& -\log Z_\alpha^G(p_{\theta_1}:p_{\theta_2}),\\
&=& J_F(p_{\theta_1}:p_{\theta_2}).
\end{eqnarray}

In the literature, the Bhattacharyya distance $\Bhat_\alpha(p:q)$ is often defined as our reverse Bhattacharyya distance:
\begin{equation}\label{eq:bhat2}
\Bhat_\alpha(p:q) := -\log \int_\calX p^\alpha q^{1-\alpha}  \dmu = B_\alpha(q:p). 
\end{equation}

\begin{corollary}
The JS-symmetrization of the reverse Kullback-Leibler divergence between densities of the same exponential family amount to calculate a Jensen/Burbea-Rao divergence between the corresponding natural parameters.
\end{corollary}


Let us report one numerical example:
Consider $\alpha=\frac{1}{2}$, and
the source parameters $\lambda_v^1=\mu_1=\pvec{0}{0}, \lambda_M^1=\Sigma_1=\pmat{1}{0}{0}{1}$ and
$\lambda_v^2=\mu_2=\pvec{1}{2}, \lambda_M^2=\Sigma_2=\pmat{1}{-1}{-1}{2}$.
The corresponding natural parameters are  $\theta_v^1=\pvec{0}{0}, \theta_M^1=\pmat{\frac{1}{2}}{0}{0}{\frac{1}{2}}$ and
 $\theta_v^2=\pvec{4}{3}, \theta_M^2=\pmat{1}{\frac{1}{2}}{\frac{1}{2}}{\frac{1}{2}}$, and the
dual expectation parameters are  $\eta_v^1=\pvec{0}{0}, \eta_M^1=\pmat{-1}{0}{0}{-1}$ and $\eta_v^2=\pvec{1}{2}, \eta_M^2=\pmat{-2}{-1}{-1}{6}$.
The interpolated parameter is 
	 $\theta_v^{\alpha}=\pvec{2}{\frac{3}{2}}, \theta_M^{\alpha}=\pmat{\frac{3}{4}}{\frac{1}{4}}{\frac{1}{4}}{\frac{1}{2}}$, or
	equivalently
	 $\eta_v^{\alpha}=\pvec{1}{1}, \eta_M^{\alpha}=\pmat{\frac{9}{5}}{\frac{3}{5}}{\frac{3}{5}}{\frac{11}{5}}$ or
	  $\lambda_v^{\alpha}=\pvec{1}{1}, \lambda_M^{\alpha}=\pmat{\frac{4}{5}}{-\frac{2}{5}}{-\frac{2}{5}}{\frac{6}{5}}$.
We find that geometric Jensen-Shannon KL divergence is $\simeq 1.26343$
	and the geometric Jensen-Shannon reverse KL divergence is $\simeq 0.86157$.

	In \S\ref{sec:BhatGJS}, we further show that the Bhattacharyya distance $B_\alpha$ between two densities is the negative of the log-normalized of an exponential family induced by the geometric mixtures of the densities.

\subsubsection{Applications to $k$-means clustering}\label{sec:kmpp}

Let $P=\{p_1,\ldots, p_n\}$ denote a point set, and $C=\{c_1,\ldots, c_k\}$ denote a set of $k$ (cluster) centers.
The generalized $k$-means objective~\cite{Bregman-2005} with respect to a distance $D$ is defined by:
\begin{equation}
E_D(P,C) = \frac{1}{n}\sum_{i=1}^n \min_{j\in\{1,\ldots,k\}} D(p_i:c_j).
\end{equation}
By defining the distance $D(p,C)=\min_{j\in\{1,\ldots,k\}} D(p:c_j)$ of a point to a set of points,
we can rewrite compactly the objective function as $E_D(P,C) = \frac{1}{n}\sum_{i=1}^n D(p_i,C)$.
Denote by $E_D^*(P,k)$ the minimum objective loss for a set of $k=|C|$ clusters: $E_D^*(P,k)=\min_{|C|=k} E_D(P,C)$.
It is NP-hard~\cite{BregmanClustering-2014} to compute $E_D^*(P,k)$ when $k>1$ and the dimension $d>1$.
The most common heuristic is Lloyd's batched $k$-means~\cite{Bregman-2005} that yields a local minimum.

The performance of the {\em probabilistic $k$-means++ initialization}~\cite{kmpp-2007} has been extended to arbitrary distances in~\cite{clustering-2014} as follows:

\begin{theorem}[Generalized $k$-means++ performance, \cite{tJ-2015}]\label{thm:kmeansplus}
Let $\kappa_1$ and $\kappa_2$ be two constants such that $\kappa_1$ defines the
quasi-triangular inequality property:
\begin{equation}
D(x:z) \leq \kappa_1 \left(D(x:y)+D(y:z)\right),\quad\forall{}x,y,z\in\Delta^d,
\end{equation}
and $\kappa_2$ handles the symmetry inequality:
\begin{equation}
D(x:y)\leq \kappa_2 D(y:x),\quad\forall x,y\in\Delta^d.
\end{equation}
Then the generalized $k$-means++ seeding guarantees with high probability a configuration $C$ of cluster centers such that:
\begin{equation}\label{eq:kmperf}
E_D(P,C)\leq 2\kappa_1^2(1+\kappa_2)(2+\log k) E_D^*(P,k).
\end{equation}
\end{theorem}

To bound the constants $\kappa_1$ and $\kappa_2$, we rewrite the generalized Jensen-Shannon divergences using quadratic form expressions:
That is, using a squared Mahalanobis distance:
\begin{equation}
D_Q(p:q)=\sqrt{(p-q)^\top Q (p-q)},
\end{equation}
for a positive-definite matrix $Q\succ 0$.
Since the Bregman divergence can be interpreted as the tail of a first-order Taylor expansion, we have:
\begin{equation}
B_F(\theta_1:\theta_2)= \frac{1}{2} (\theta_1-\theta_2)^\top \nabla^2 F(\xi) (\theta_1-\theta_2),
\end{equation}
for $\xi\in\Theta$ (open convex).
Similarly, the Jensen divergence can be interpreted as a Jensen-Bregman divergence, and thus we have
\begin{equation}
J_F(\theta_1:\theta_2) \frac{1}{2} (\theta_1-\theta_2)^\top \nabla^2 F(\xi') (\theta_1-\theta_2),
\end{equation}
for $\xi'\in\Theta$.
More precisely, for a prescribed point set $\{\theta_1, \ldots, \theta_n\}$, we have $\xi,\xi'\in \CH(\{\theta_1, \ldots, \theta_n\})$, where $\CH$ denotes the closed convex hull.
We can therefore upper bound $\kappa_1$ and $\kappa_2$ using the ratio 
$\frac{\max_{\theta\in \CH(\{\theta_1, \ldots, \theta_n\})} \|\nabla^2 F(\theta)\| }{\max_{\theta\in \CH(\{\theta_1, \ldots, \theta_n\})} \|\nabla^2 F(\theta)\|}$.  
See~\cite{Bregman-2010} for further details.

A centroid for a set of parameters $\theta_1, \ldots, \theta_n$ is defined as the minimizer of the functional
\begin{equation}
E_D(\theta)= \frac{1}{n}\sum_i D(\theta_i:\theta).
\end{equation}

In particular, the {\em symmetrized Bregman centroids} have been studied in~\cite{SBD-2009} (for $\JS^{G_\alpha}$), and the {\em Jensen centroids} (for $\JS^{G_\alpha}_*$) have been investigated in~\cite{BR-2011} using the convex-concave iterative procedure.

\subsection{Properties of the $M$-Jensen--Shannon divergence}\label{sec:propMJS}

The $M$-Jensen--Shannon divergence is a $M$-divergence induced by the Kullback--Leibler base divergence:
$$
\JS^M(p,q)=\frac{1}{2}\left(\KL\left(p:(pq)^M\right) + \KL\left(q:(pq)^M\right)\right),
$$
where $(pq)^M$ denotes the $M$-mixture of densities $p(x)$ and $q(x)$.
The ordinary Jensen--Shannon divergence is recovered for the arithmetic mean $M=A$ with $(pq)^A(x)=\frac{p(x)+q(x)}{2}$:
$$
\JS^A(p,q)=\frac{1}{2}\left(\KL\left(p:(pq)^A\right) + \KL\left(q:(pq)^A\right)\right)=\JS(p,q).
$$
Let $K^M(p:q)=\KL\left(p:(pq)^M\right)$.
For symmetric abstract means $M(a,b)=M(b,a)$ (and $(qp)^M=(pq)^M$), we have~\cite{SymJensen-2010}:
$$
\JS^M(p,q)=\frac{1}{2}\left(K^M(p:q)+K^M(q:p)\right).
$$
When the densities are unnormalized, the extended KLD is defined as 
$$
\KL(\tilde{p}:\tilde{q})=\int \left( \tilde p(x)\log\frac{\tilde p(x)}{\tilde q(x)}+\tilde q(x)-\tilde p(x)\right)\dmu(x),
$$
and the generalized Jensen-Shannon divergence can be extended accordingly.

The $M$-Jensen--Shannon divergence enjoys the following decomposition:
\begin{property}
We have the following identity:
\begin{equation}
\JS^M(p,q) = \JS(p,q) + \KL\left(\frac{p+q}{2}:(pq)^M\right).
\end{equation}
\end{property}

\begin{proof}
Recall that the KLD is the difference between the cross-entropy and the entropy:
$$
\KL(p:q)=h_\times(p:q)-h(p),
$$
with $h_\times(p:q)\geq h(p)$ (and hence $\KL(p:q)\geq 0$).

Thus we have
\begin{eqnarray*}
\JS^M(p,q) &=& \frac{1}{2}\left(\KL\left(p:(pq)^M\right) + \KL\left(q:(pq)^M\right)\right),\\
&=& \frac{1}{2} h_\times\left(p:(pq)^M\right) - \frac{1}{2} h(p) + \frac{1}{2} h_\times\left(q:(pq)^M\right) - \frac{1}{2} h(q),\\
&=& h_\times\left(\frac{p+q}{2}:(pq)^M\right)-\frac{h(p)+h(q)}{2}
\end{eqnarray*}
since $\alpha h_\times(p_1:q)+(1-\alpha )h_\times(p_2:q)=h_\times(\alpha p_1+(1-\alpha p_2):q)$ in general.
Therefore we have
\begin{eqnarray*}
\JS^M(p,q) &=&  h_\times\left(\frac{p+q}{2}:(pq)^M\right)-h\left(\frac{p+q}{2}\right)+h\left(\frac{p+q}{2}\right) -\frac{h(p)+h(q)}{2},\\ 
&=& \KL\left(\frac{p+q}{2}:(pq)^M\right)+\JS(p,q).
\end{eqnarray*}
\end{proof}

Since $\KL\left(\frac{p+q}{2}:(pq)^M\right)\geq 0$ (Gibb's inequality), it follows the following corollary:

\begin{corollary}
We have $\JS^M(p,q)\geq \JS(p,q)$.
\end{corollary}
That is the $M$-JS divergence always upper bounds the ordinary Jensen--Shannon divergence.

Notice that we could have also derived this result from the following identity: 
$$
\JS^M(p,q) = h_\times\left(\frac{p+q}{2}:(pq)^M\right)-\frac{h(p)+h(q)}{2}
$$
since the cross-entropy $h_\times(p,q)$ always upper bounds the entropy $h(p)$.
Therefore, $h_\times\left(\frac{p+q}{2}:(pq)^M\right)\geq h\left(\frac{p+q}{2}\right)$, and we have:
\begin{eqnarray}
\JS^M(p,q) &=&  h_\times\left(\frac{p+q}{2}:(pq)^M\right)-\frac{h(p)+h(q)}{2},\\
&\geq& h\left(\frac{p+q}{2}\right)-\frac{h(p)+h(q)}{2}=\JS(p,q).
\end{eqnarray}

When an abstract weighted mean $M_\alpha$ dominates another weighted abstract mean $N_\alpha$ (i.e., $M_\alpha\geq N_\alpha$), we have 
$Z^M_\alpha(p,q)\geq Z^N_\alpha(p,q)$.

To compare $\JS^{M_\alpha}(p,q)$ with $\JS^{N_\alpha}(p,q)$, we have to decide the sign of $\JS^{M_\alpha}(p,q)-\JS^{N_\alpha}(p,q)$ which amounts to calculate the sign of the following expression when $\alpha=\frac{1}{2}$:
$$
\KL\left(\frac{p+q}{2}:(pq)^M\right)-\KL\left(\frac{p+q}{2}:(pq)^N\right)=
\int \frac{p(x)+q(x)}{2} \log \frac{N(p(x),q(x))}{M(p(x),q(x)}  \frac{Z^M_{\frac{1}{2}}(p,q)}{Z^N_{\frac{1}{2}}(p,q)} \dmu(x).
$$ 
Thus we have
$$
\KL\left(\frac{p+q}{2}:(pq)^M\right)-\KL\left(\frac{p+q}{2}:(pq)^N\right)=
\log \frac{Z^M_{\frac{1}{2}}(p,q)}{Z^N_{\frac{1}{2}}(p,q)} + \int \frac{p(x)+q(x)}{2} \log \frac{N(p(x),q(x))}{M(p(x),q(x)} \dmu(x),
$$
where $\log \frac{Z^M_{\frac{1}{2}}(p,q)}{Z^N_{\frac{1}{2}}(p,q)} \geq 0$ (when $M\geq N$) and
$\int \frac{p(x)+q(x)}{2} \log \frac{N(p(x),q(x))}{M(p(x),q(x)} \dmu(x)\leq 0$ (when $M\geq N$).

When $M=G$, the geometric Jensen-Shannon divergence can be rewritten using familiar divergences using the fact that $(pq)^G(x)=\frac{\sqrt{p(x)q(x)}}{Z^G(p,q)}$ as follows:

\begin{eqnarray*}
\JS^G(p,q) &:=& \frac{1}{2} \left(\KL(p:(pq)^G)+\KL(q:(pq)^G)\right),\\
&=& \frac{1}{2} \left(\int \left(p(x)\log\frac{p(x)\, Z^G(p,q)}{\sqrt{p(x)q(x)}}+ q(x)\log\frac{q(x)\, Z^G(p,q)}{\sqrt{p(x)q(x)}}\right)\dmu(x) \right),\\
&=& \frac{1}{2}\left(\int \left(p(x)+q(x)\right)\log Z^G(p,q)\dmu(x)+\frac{1}{2}\KL(p:q)+\frac{1}{2}\KL(q:p)  \right),\\
&=& \log Z^G(p,q) + \frac{1}{4} J(p,q),\\
&=& \frac{1}{4} J(p,q) - B(p,q),
\end{eqnarray*}
where $J(p,q)$ is Jeffreys' divergence and $B(p,q)$ is the Bhattacharyya distance: $B(p,q)=-\log\int\sqrt{p(x)q(x)}\dmu(x)$.

\begin{proposition}
We have $\JS^G(p,q)=\frac{1}{4} J(p,q) - B(p,q)$.
Therefore we get the inequality $J(p,q)\geq 4\, B(p,q)$ since $\JS^G(p,q)\geq 0$.
\end{proposition}

When $p=p_{\theta_1}$ and $q=p_{\theta_2}$ belongs to a same exponential family with cumulant function $F(\theta)$, we conclude that

\begin{equation}
\JS^G(p_{\theta_1},p_{\theta_2}) =
\underbrace{\frac{1}{4}(\theta_2-\theta_1)^\top(\nabla F(\theta_2)-\nabla F(\theta_1))}_{\frac{1}{4} J(p_{\theta_1},p_{\theta_2})}-\underbrace{\left(
\frac{F(\theta_1)+F(\theta_2)}{2}-F\left(\frac{\theta_1+\theta_2}{2}\right)
\right)}_{B(p_{\theta_1},p_{\theta_2})=J_F(\theta_1,\theta_2)}.
\end{equation}

\subsection{The skewed Bhattacharyya distance interpreted as a geometric Jensen-Shannon symmetrization ($G$-$\JS$)\label{sec:BhatGJS}}

Let $\KL^*(p:q):=\KL(q:p)$ denote the reverse Kullback-Leibler divergence (rKL divergence for short).
The {\em $*$-reference duality} is an involution: $(D^*)^*(p:q)=D(p:q)$.

The geometric Jensen-Shannon symmetrization~\cite{vskewJS-2020} of the reverse Kullback-Leibler divergence is defined by
\begin{equation}
\JS_{\KL^*}^{G_\alpha}(p:q) := (1-\alpha)\KL^*(p:(pq)_\alpha^G)+\alpha\KL^*(q:(pq)_\alpha^G),
\end{equation}
where $G_\alpha(x,y)=x^{1-\alpha}y^\alpha$ denotes the geometric weighted mean for $x>0$ and $y>0$, and 
\begin{equation}
(pq)_\alpha^G(x) := \frac{G_\alpha(p(x),q(x))}{Z_\alpha^G(p:q)}, 
\end{equation}
is the geometric mixture with normalizing coefficient:
\begin{equation}
Z_\alpha^G(p:q)=\int G_\alpha(p(x),q(x))\dmu(x).
\end{equation}

Let us observe that following identity:

\begin{BoxedProposition}
The skewed Bhattacharyya distance is a geometric Jensen-Shannon divergence with respect to the reverse Kullback-Leibler divergence:
$$
\JS_{\KL^*}^{G_\alpha}(p:q)=B_\alpha(p:q).
$$
\end{BoxedProposition}

Proof:
\begin{eqnarray}
\JS_{\KL^*}^{G_\alpha}(p:q) &=& \int\left((1-\alpha)(pq)_\alpha^G\log\frac{(pq)_\alpha^G}{p}+\alpha (pq)_\alpha^G\log\frac{(pq)_\alpha^G}{q} \right)\dmu,\\
&=& \int\left((pq)_\alpha^G\log\frac{(pq)_\alpha^G}{p^{1-\alpha}q^\alpha}\right)\dmu,\\
&=& \int (pq)_\alpha^G\log \frac{1}{Z_\alpha^G(p:q)}\dmu,\\
&=& -\log Z_\alpha^G(p:q)\int (pq)_\alpha^G\dmu,\\
&=& B_\alpha(p:q),
\end{eqnarray}
since $Z_\alpha^G(p:q)=\int p^{1-\alpha}q^\alpha\dmu$ and $(pq)_\alpha^G=p^{1-\alpha}q^\alpha/Z_\alpha^G(p:q)$.

The Bhattacharyya distance can also be interpreted as {\em the negative the log-normalizer of the 1D exponential family} induced by the two  {\em distinct} distributions $p$ and $q$.
Indeed, consider the family of geometric mixtures
$$
\mathcal{E}(p,q):=\left\{ p_\lambda(x):= \frac{p^{1-\lambda}(x)q^\lambda(x)}{Z^G_\lambda(p,q)},\quad \lambda\in (0,1)\right\}.
$$
This family describes an {\em Hellinger arc}~\cite{Gzyl-1995,QuantumChernoff-2009} (also called a {\em Bhattacharyya arc}). 
Let $p_0(x):=p(x)$ and $p_1(x):=q(x)$.
The Bhattacharyya arc is a 1D natural exponential family  (i.e., an exponential family of order $1$):

\begin{eqnarray}
p_\lambda(x) &=& \frac{p_0^{1-\lambda}(x)p_1^\lambda(x)}{Z^G_\lambda(p,q)},\\
&=& p_0(x) \exp\left(\lambda \log\left(\frac{p_1(x)}{p_0(x)}\right) -\log Z^G_\lambda(p,q)\right),\\
&=& \exp\left(\lambda t(x)-F_{pq}(\lambda)+k(x)\right),
\end{eqnarray}
with sufficient statistics $t(x)$ the logarithm of {\em likelihood ratio}\footnote{Hence, Gr\"unwald~\cite{MDL-Grunwald-2007} called this Hellinger arc a {\em likelihood ratio exponential family}.} (i.e., the ratio of the densities $\frac{p_1(x)}{p_0(x)}$):
\begin{equation}
t(x):=\log\left(\frac{p_1(x)}{p_0(x)}\right),
\end{equation}
auxiliary carrier term $k(x)=\log p(x)$ with respect to $\mu$ (e.g., counting or Lebesgue positive measure), 
natural parameter $\theta=\lambda$, and log-normalizer:
\begin{eqnarray}
F_{pq}(\lambda) = F_{pq}(\theta) &:=& \log\left(Z^G_\lambda(p,q)\right),\\
&=& \log \left(\int_{\calX} {p^{1-\lambda}(x)q^\lambda(x)}\dmu(x)\right),\\
&=:& -B_\lambda[p:q] ,
\end{eqnarray}
where $B_\alpha$ is the skewed $\alpha$-Bhattacharyya distance also called $\alpha$-Chernoff distance.

Since the log-normalizer $F_{pq}$ of an exponential family is strictly convex and real analytic $C^\omega$, we deduce that $B_\alpha[p:q]$ is strictly concave and real analytic for $\alpha\in (0,1)$. Moreover, since $B_0[p:q]=B_1[p:q]=0$, there exists are unique value $\alpha^*$ maximizing 
$B_\alpha[p:q]$:
\begin{equation}
\alpha^*=\arg\max_{\alpha\in [0,1]} B_\alpha[p:q].
\end{equation}

The maximal skewed Bhattacharyya distance is called the {\em Chernoff information}~\cite{Chernoff-2011,Chernoff-2013}:
\begin{equation}
D^{\Chernoff}_\alpha[p:q] := B_{\alpha^*}[p:q].
\end{equation}
The value $\alpha^*$ corresponds to the error exponent in Bayesian asymptotic hypothesis testing~\cite{Chernoff-2013,GenBhat-2014}.
Note that since $F_{pq}(\lambda)<\infty$ for $\lambda\in (0,1)$, we have $B_\alpha[p:q]<\infty$ for $\alpha\in (0,1)$.
Moreover, the moment parameterization of a 1D exponential family~\cite{EF-2009} is $\eta(\theta)=E_{p_\theta}[t(x)]=F'(\theta)$.
Thus it follows that 
\begin{equation}
F'_{pq}(\lambda)=E_{p_\lambda}[t(x)]=E_{p_\lambda}\left[\log\left(\frac{p_1(x)}{p_0(x)}\right)\right].
\end{equation}
The optimal exponent $\alpha^*$ for the Chernoff information is therefore found for $B'_\alpha(p:q)=-F'_{pq}(\alpha)=0$:
\begin{eqnarray}
E_{p_\lambda}\left[\log\left(\frac{p_0(x)}{p_1(x)}\right)\right] &=& 0,\\
\int p_\lambda(x)\log\left(\frac{p_0(x)}{p_1(x)}\right)\dmu(x) &=& 0,\\
\int p_\lambda(x)\log\left(\frac{p_0(x)p_\lambda(x)}{p_\lambda(x)p_1(x)}\right)\dmu(x) &=& 0,\\
\int p_\lambda(x)\log\left(\frac{p_0(x)}{p_\lambda(x)}\right)\dmu(x) + 
\int p_\lambda(x)\log\left(\frac{p_\lambda(x)}{p_1(x)}\right)\dmu(x) &=& 0,\\
\KL(p_\lambda:p_1)-\KL(p_\lambda:p_0)&=&0.
\end{eqnarray}
Thus at the optimal exponent $\alpha^*$, we have $\KL(p_{\alpha^*}:p_1)=\KL(p_{\alpha^*}:p_0)$.

We summarize the result:
\begin{BoxedProposition}
The skewed Bhattacharyya distance $B_\alpha$ is strictly concave on $(0,1)$ and real analytic: 
$B_\alpha$ admits a unique maximum $\alpha^*\in(0,1)$ called the Chernoff information so that $\KL(p_{\alpha^*}:p_1)=\KL(p_{\alpha^*}:p_0)$.
\end{BoxedProposition}

Since the reverse KL divergence between two densities amount to a Bregman divergence between the corresponding natural parameters for the Bregman generator set to the cumulant function~\cite{EIG-2018,EIG-2020}, we have:

\begin{equation}
\KL^*((pq)_{\lambda_1}^G:(pq)_{\lambda_2}^G)=B_{F_{pq}}(\lambda_1:\lambda_2)=\KL((pq)_{\lambda_2}^G:(pq)_{\lambda_1}^G),
\end{equation} 
where $B_{F_{pq}}$ is the Bregman divergence corresponding to the univariate generator $F_{pq}(\theta)=-B_\theta(p:q)$.
Therefore we check the following identity:
\begin{equation}
\KL^*((pq)_{\lambda_1}^G:(pq)_{\lambda_2}^G)=\log\left(\frac{\int p^{1-\beta}q^\beta\dmu}{\int p^{1-\beta}q^\beta\dmu}\right)-(\beta-\alpha)\left(\KL(p_\alpha:p_1)-\KL(p_\alpha:p_0)\right).
\end{equation}

\subsection{The harmonic Jensen-Shannon divergence ($H$-$\JS$)\label{sec:HJS}}

The {\em principle} to get closed-form formula for generalized Jensen-Shannon divergences between distributions belonging to a parametric family 
$\calP_\Theta = \{p_\theta \st \theta\in\Theta\}$ consists in finding an abstract mean $M$ such that 
the $M$-mixture $(p_{\theta_1}p_{\theta_2})_\alpha^M$ belongs to the family $\calP_\Theta$.
In particular, when $\Theta$ is a convex domain, we seek a mean $M$ such that $(p_{\theta_1}p_{\theta_2})_\alpha^M = p_{(\theta_1\theta_2)_\alpha}$ 
with $(\theta_1\theta_2)_\alpha\in\Theta$.

Let us consider the {\em weighted harmonic mean}~\cite{Convexity-2006} (induced by the harmonic mean) $H$:
\begin{equation}
H_\alpha(x,y)  \eqdef \frac{1}{(1-\alpha)\frac{1}{x}+\alpha \frac{1}{y}}  =\frac{xy}{(1-\alpha)y+\alpha x} = \frac{xy}{(xy)_{1-\alpha}},\quad \alpha\in [0,1].
\end{equation}

The harmonic mean is a quasi-arithmetic mean $H_\alpha(x,y)=M^h_\alpha(x,y)$ obtained for the monotone (decreasing) function $h(u)=\frac{1}{u}$ 
(or equivalently for the increasing monotone function $h(u)=-\frac{1}{u}$).

This harmonic mean is well-suited for the {\em scale family} $\calC$ of Cauchy probability distributions~\cite{CauchyEntropy-2020} (also called Lorentzian distributions):
\begin{equation}
\calC_{\Gamma} \eqdef \left\{\ p_\gamma(x)= \frac{1}{\gamma}p_{\std}\left(\frac{x}{\gamma}\right) = \frac{\gamma}{\pi(\gamma^2+x^2)} \st \gamma\in\Gamma=(0,\infty) \right\},
\end{equation}
where $\gamma$ denotes the scale and $p_{\std}(x)=\frac{1}{\pi (1+x^2)}$ the {\em standard Cauchy distribution}.

Using a computer algebra system\footnote{We use Maxima: \url{http://maxima.sourceforge.net/}} (see Appendix~\ref{sec:cauchymax}), 
we find that
\begin{equation}
(p_{\gamma_1}p_{\gamma_2})^H_{\frac{1}{2}}(x) =
 \frac{H_\alpha(p_{\gamma_1}(x):p_{\gamma_2}(x))}{Z_\alpha^H(\gamma_1,\gamma_2)} =  
 p_{(\gamma_1\gamma_2)_\alpha} 
\end{equation}
where the normalizing coefficient is
\begin{equation}\label{eq:ZH}
Z_\alpha^H(\gamma_1,\gamma_2) \eqdef \sqrt{  \frac{\gamma_1\gamma_2}{(\gamma_1\gamma_2)_\alpha (\gamma_1\gamma_2)_{1-\alpha}}}  
= \sqrt{  \frac{\gamma_1\gamma_2}{(\gamma_1\gamma_2)_\alpha (\gamma_2\gamma_1)_{\alpha}}},
\end{equation}
since we have $(\gamma_1\gamma_2)_{1-\alpha}=(\gamma_2\gamma_1)_{\alpha}$.


The $H$-Jensen-Shannon symmetrization of a distance $D$ between  distributions writes as:
\begin{equation}
\JS_D^{H_\alpha}(p:q) = (1-\alpha) D(p:(pq)^H_\alpha) + \alpha D(q:(pq)^H_\alpha),
\end{equation}
where $H_\alpha$ denote the weighted harmonic mean.
When $D$ is available in closed form for distributions belonging to the scale Cauchy distributions, so is $\JS_D^{H_\alpha}(p:q)$.

For example, consider the KL divergence formula between two scale Cauchy distributions:\footnote{The formula initially reported in~\cite{KLCauchy-1997}  has been corrected by the authors.} 
\begin{equation}\label{eq:klcauchy}
\KL(p_{\gamma_1}:p_{\gamma_2}) =  2\log \frac{A(\gamma_1,\gamma_2)}{G(\gamma_1,\gamma_2)} = 2\log \frac{\gamma_1+\gamma_2}{2\sqrt{\gamma_1\gamma_2}},
\end{equation}
where $A$ and $G$ denote the arithmetic and geometric means, respectively. 
Since $A\geq G$ (and $\frac{A}{G}\geq 1$), it follows that $\KL(p_{\gamma_1}:p_{\gamma_2})\geq 0$.
Notice that the KL divergence is symmetric\footnote{For exponential families, the KL divergence is symmetric only for the location Gaussian family (since the only symmetric Bregman divergences are the squared Mahalanobis distances~\cite{BVD-2010}).} for Cauchy scale distributions.
The cross-entropy between scale Cauchy distributions is 
$h^\times(p_{\gamma_1}:p_{\gamma_2}) = \log \pi\frac{(\gamma_1+\gamma_2)^2}{\gamma_2}$, and the differential entropy is
$h(p_{\gamma})=h^\times(p_{\gamma}:p_{\gamma})=\log 4\pi\gamma$.

Then the $H$-JS divergence between $p=p_{\gamma_1}$ and $q=p_{\gamma_2}$ is:

\begin{eqnarray}
\JS^{H}(p:q) &=& \frac{1}{2} \left( \KL\left(p:(pq)^H_{\frac{1}{2}}\right) + \KL\left(q:(pq)^H_{\frac{1}{2}}\right) \right),\\
\JS^{H}(p_{\gamma_1}:p_{\gamma_2}) &=& \frac{1}{2} \left( \KL\left(p_{\gamma_1}: p_{\frac{\gamma_1+\gamma_2}{2}}\right) + \KL\left(p_{\gamma_2}: p_{\frac{\gamma_1+\gamma_2}{2}}\right) \right),\\
&=& \log \left( \frac{(3\gamma_1+\gamma_2)(3\gamma_2+\gamma_1)}{8\sqrt{\gamma_1\gamma_2} (\gamma_1+\gamma_2)}\right).
\end{eqnarray}

We check that when $\gamma_1=\gamma_2=\gamma$, we have $\JS^{H_\alpha}(p_\gamma:p_\gamma)=0$.
Notice that we could use the more generic KL divergence formula between two location-scale Cauchy distributions $p_{l_1,\gamma_1}$ and $p_{l_2,\gamma_2}$ (with respective location $l_1$ and $l_2$):
\begin{equation}
\KL(p_{l_1,\gamma_1}:p_{l_2,\gamma_2})=\log\frac{(\gamma_1+\gamma_2)^2+(l_1-l_2)^2}{4\gamma_1\gamma_2}.
\end{equation}

\begin{theorem}[Harmonic JSD between scale Cauchy distributions.]\label{thm:hjsd} 
The harmonic Jensen-Shannon divergence between two scale Cauchy distributions $p_{\gamma_1}$ and $p_{\gamma_2}$ is \\
$\JS^{H}(p_{\gamma_1}:p_{\gamma_2}) =  \log \frac{(3\gamma_1+\gamma_2)(3\gamma_2+\gamma_1)}{8\sqrt{\gamma_1\gamma_2} (\gamma_1+\gamma_2)}$.
\end{theorem}

Let us report some numerical examples:
Consider $p_{\gamma_1}=0.1$ and $p_{\gamma_1}=0.5$, we find that $\JS^{H}(p_{\gamma_1}:p_{\gamma_2})\simeq 0.176$.
When $p_{\gamma_1}=0.2$ and $p_{\gamma_1}=0.8$, we find that $\JS^{H}(p_{\gamma_1}:p_{\gamma_2})\simeq 0.129$.

Notice that KL formula is scale-invariant and this property holds for any scale family:
\begin{lemma}
The Kullback-Leibler divergence between two distributions $p_{s_1}$ and $p_{s_2}$ belonging to 
the same scale family $\{p_s(x)=\frac{1}{s}p(\frac{x}{s})\}_{s\in (0,\infty)}$ with standard density $p$ is scale-invariant:
$\KL(p_{\lambda s_1}: p_{\lambda s_2})=\KL(p_{s_1}: p_{s_2})=\KL(p: p_{\frac{s_2}{s_1}})=\KL(p_{\frac{s_1}{s_2}}: p)$ for any $\lambda>0$.
\end{lemma}

A direct proof follows from a change of variable in the KL integral with $y=\frac{x}{\lambda}$ and $\mathrm{d}x=\lambda\mathrm{d}y$.
Note that although the KLD between scale Cauchy distributions is symmetric, it is not the case for all scale families:
For example, the Rayleigh distributions form a scale family with the KLD amounting to compute a Bregman asymmetric Itakura-Saito divergence between parameters~\cite{EF-2009}.

Instead of the KLD, we can choose the total variation distance for which a formula has been reported in~\cite{GenBhat-2014} between two Cauchy distributions.
Notice that the Cauchy distributions are alpha-stable distributions for $\alpha=1$ and $q$ gaussian distributions for $q=2$ (\cite{Naudts-2011}, p. 104).
A closed-form formula for the divergence between two $q$-Gaussians is given in~\cite{Naudts-2011} when $q<2$.
The definite integral $h_q(p)=\int_{-\infty}^{+\infty} p(x)^q \dmu$ is available in closed-form for Cauchy distributions.
When $q=2$, we have $h_2(p_\gamma)=\frac{1}{2\pi\gamma}$.

This example further extends to power means and Student $t$-distributions:
We refer to~\cite{GenBhat-2014} for yet other illustrative examples considering the family of Pearson type VII distributions and central multivariate $t$-distributions which use the power means (quasi-arithmetic means $M^h$ induced by $h(u)=u^\alpha$ for $\alpha>0$) for defining mixtures.

Table~\ref{tab:Z} summarizes the various examples introduced in the paper.

\begin{table}
\centering
\begin{tabular}{llll}
$\JS^{M_\alpha}$ & mean $M$ & parametric family & $Z_\alpha^M(p:q)$\\ \hline
$\JS^{A_\alpha}$ & arithmetic $A$ & mixture family & $Z_\alpha^M(\theta_1:\theta_2)=1$\\
$\JS^{G_\alpha}$ & geometric $G$ & exponential family & $Z_\alpha^G(\theta_1:\theta_2)=\exp(-J_F^\alpha(\theta_1:\theta_2))$\\
$\JS^{H_\alpha}$ & harmonic $H$ & Cauchy scale family & $Z_\alpha^H(\theta_1:\theta_2)=\sqrt{  \frac{\theta_1\theta_2}{(\theta_1\theta_2)_\alpha (\theta_1\theta_2)_{1-\alpha}}}$\\ \hline
\end{tabular}

\caption{Summary of the weighted means $M$ chosen according to the parametric family in order to ensure that the family is closed under $M$-mixturing: $(p_{\theta_1}p_{\theta_2})_\alpha^M=p_{(\theta_1\theta_2)_\alpha}$.  \label{tab:Z}}
\end{table}

\subsection{The $M$-Jensen-Shannon matrix distances\label{sec:MatrixJS}}

In this section, we consider distances between matrices which play an important role in quantum computing~\cite{briet2009properties,matrixJS-2014}. We refer to~\cite{JBLogDet-2013} for the matrix Jensen-Bregman logdet divergence.
The {\em Hellinger distance} can be interpreted as the difference of an arithmetic mean $A$ and a geometric mean $G$:

\begin{equation}
D_H(p,q) = \sqrt{1-\int_\calX \sqrt{p(x)}\sqrt{q(x)}\dmu(x)} = \sqrt{ \int_\calX  (A(p(x),q(x))-G(p(x),q(x))) \dmu(x)}.
\end{equation}

Notice that since $A\geq G$, we have $D_H(p,q)\geq 0$.
The scaled and squared Hellinger distance is an $\alpha$-divergence $I_\alpha$ for $\alpha=0$.
Recall that the $\alpha$-divergence can be interpreted as the difference of a weighted arithmetic minus a weighted geometry mean.

In general, if a mean $M_1$ dominates a mean $M_2$, we may define the distance as
\begin{equation}
D_{M_1,M_2}(p,q) = \int_\calX  \left( M_1(p,q)-M_2(p,q) \right) \dmu(x).
\end{equation}

However, when considering matrices~\cite{Bhatia-2018}, there is {\em not} a unique definition of a geometric matrix mean, and thus
we have different notions of {\em matrix Hellinger distances}~\cite{Bhatia-2018}, some of them are divergences (i.e., smooth distances defining a dualistic structure in information geometry).

We define the {\em matrix $M$-Jensen-Shannon divergence} for a matrix divergence $D$ as follows:

\begin{equation}
\JS_D^M(X_1,X_2) :=  \frac{1}{2} \left( D(X_1,M(X_1,X_2)) + D(X_2,M(X_1,X_2))  \right).
\end{equation}

\section{Conclusion and perspectives}\label{sec:concl}

In this work, we introduced a generalization of the celebrated Jensen-Shannon divergence~\cite{Lin-1991}, 
termed the {\em $(M,N)$-Jensen-Shannon divergences}, based on statistical {\em $M$-mixtures} derived from generic abstract means $M$ and $N$, where the $N$ mean is used to symmetrize the  asymmetric Kullback-Leibler divergence.
This new family of  divergences includes  the ordinary Jensen-Shannon divergence when both $M$ and $N$ are set to the arithmetic mean.
This process can be extended to any base divergence $D$ to obtain its JS-symmetrization.
We reported closed-form expressions of the $M$ Jensen-Shannon divergences for mixture families and exponential families in information geometry 
by choosing the arithmetic and geometric weighted mean, respectively.
The {\em $\alpha$-skewed  geometric Jensen-Shannon divergence} ($G$-JSD for short) between densities $p_{\theta_1}$ and $p_{\theta_2}$ of the same exponential family with cumulant function $F$ is 
$$
\JS_{\KL}^{G_\alpha}[p_{\theta_1}:p_{\theta_2}] = \JS_{{B_F}^*}^{A_\alpha}(\theta_1:\theta_2).
$$
Here, we used the bracket notation to emphasize the fact that the statistical distance $\JS_{\KL}^{G_\alpha}$ is between densities, and the parenthesis notation to emphasize that the  distance $\JS_{{B_F}^*}^{A_\alpha}$ is between (vector) parameters.
We also have $\JS_{\KL^*}^{G_\alpha}[p_{\theta_1}:p_{\theta_2}]=J_F^\alpha(\theta_1:\theta_2)$.
We reported how to get a closed-form formula for the harmonic Jensen-Shannon divergence  of Cauchy scale distributions by taking harmonic mixtures.

We defined the {\em skew $N$-Jeffreys symmetrization} for an arbitrary distance $D$ and scalar $\beta\in[0,1]$:
\begin{equation}
J^{N_\beta}_D(p_1:p_{2}) = N_\beta(D(p_{1}:p_{2}),D(p_{_2}:p_{1})),
\end{equation}
and the {\em skew $(M,N)$-JS symmetrization} of an arbitrary distance $D$:
\begin{equation}
\JS_D^{M_\alpha,N_\beta}(p_{1}:p_{2}) = N_\beta(D(p_{1}, (p_{1}p_{2})_\alpha^M) , 
D(p_2, (p_{1}p_{2})_\alpha^M)).
\end{equation}

The geometric Jensen-Shannon divergence has recently found applications in machine learning~\cite{GJS-Sutter-2020,GJS-Deasy-2020}.
Finally, let us mention that the Jensen-Shannon divergence has further been extended recently to a skew vector parameter in~\cite{vskewJS-2020} instead of an ordinary scalar parameter, and generalized from the variational point of view to yield extensions of the Sibson's information radius~\cite{nielsen2021variational}.

\bibliographystyle{plain}
\bibliography{GenJSBIBV5}

\appendix

\section{Summary of distances and their notations}\label{sec:im}

\begin{supertabular}{ll}
Weighted mean & $M_\alpha$, $\alpha\in (0,1)$\\ \hline\hline
Arithmetic mean & $A_\alpha(x,y)=(1-\alpha)x+\alpha y$\\
Geometric mean   & $G_\alpha(x,y)=x^{1-\alpha}  y^\alpha$\\
Harmonic mean  & $H_\alpha(x,y)=\frac{xy}{(1-\alpha)y+\alpha x}$\\
Power mean & $P_\alpha^p(x,y)= ((1-\alpha)x^p+\alpha y^p)^{\frac{1}{p}},\quad p\in\bbR\backslash\{0\}$, $\lim_{p\rightarrow 0} P_\alpha^p=G$\\
Quasi-arithmetic mean & $M_\alpha^f(x,y)=  f^{-1}((1-\alpha)f(x)+\alpha f(y))$, $f$ strictly monotonous\\
$M$-mixture & $Z_\alpha^M(p,q) =  \int_{t\in\calX} M_\alpha(p(t),q(t)) \dmu(t)$\\
& with  $Z_\alpha^M(p,q)=  \int_{t\in\calX} M_\alpha(p(t),q(t)) \dmu(t)$\\
\\
\\
Statistical distance & $D(p:q)$\\ \hline\hline
Dual/reverse distance $D^*$ & $D^*(p:q) \eqdef  D(q:p)$\\
Kullback-Leibler divergence & $\KL(p:q)= \int p(x)\log \frac{p(x)}{q(x)} \dmu(x)$\\
reverse Kullback-Leibler divergence & $\KL^*(p:q)=\KL(q:p) =  \int  q(x) \log \frac{q(x)}{p(x)}\dmu(x)$ \\
Jeffreys divergence & $J(p;q) = \KL(p:q)+\KL(q:p) = \int (p(x)-q(x))\log \frac{p(x)}{q(x)} \dmu(x)$ \\
Resistor divergence & $R(p;q) = \frac{2\KL(p:q)\KL(q:p)}{J(p;q)}$. 
$R(p;q)=\frac{2J(p;q)}{\KL(p:q)\KL(q:p)}$\\
skew $K$-divergence & $K_\alpha(p:q) =   \int p(x)\log \frac{p(x)}{(1-\alpha)p(x)+\alpha q(x)} \dmu(x)$ \\
Jensen-Shannon divergence & $\JS(p,q) = \frac{1}{2} \left( \KL\left(p:\frac{p+q}{2}\right) +  \KL\left(q:\frac{p+q}{2}\right) \right)$ \\
skew Bhattacharrya divergence & $B_\alpha(p:q) = -\log \int_\calX p(x)^{1-\alpha} q(x)^\alpha \dmu(x)$  \\
skew Bhattacharrya divergence & $\Bhat_\alpha(p:q) = -\log \int_\calX p(x)^{\alpha} q(x)^{1-\alpha} \dmu(x)$  \\
Hellinger distance & $D_H(p,q) = \sqrt{1-\int_\calX \sqrt{p(x)}\sqrt{q(x)}\dmu(x)}$  \\
$\alpha$-divergences &  $I_\alpha({p}:{q}) =  \int \left(\alpha p(x) +(1-\alpha)q(x)-p(x)^\alpha q(x)^{1-\alpha} \right) \dmu(x), \alpha\not\in\{0,1\}$\\
&  $I_\alpha({p}:{q}) =  A_\alpha(q:p)-G_\alpha(q:p)$\\
Mahalanobis distance & $D_Q(p:q)=\sqrt{(p-q)^\top Q (p-q)}$ for a positive-definite matrix $Q\succ 0$ \\
$f$-divergence & $I_f(p:q)=\int p(x) f\left(\frac{q(x)}{p(x)}\right)\dmu(x)$, with $f(1)=f'(1)=0$\\
& $f$ strictly convex at $1$\\
reverse $f$-divergence & $I_f^*(p:q)=\int q(x) f\left(\frac{p(x)}{q(x)}\right)\dmu(x)=I_{f^\diamond}(p:q)$\\
&  for $f^\diamond(u)=u f(\frac{1}{u})$\\
J-symmetrized $f$-divergence &  $J_f(p;q)= \frac{1}{2} (I_f(p:q) + I_f(q:p)) $\\
JS-symmetrized $f$-divergence & $I_f^{\alpha}(p;q) \eqdef (1-\alpha)I_f(p:(pq)_\alpha)+\alpha I_f(q:(pq)_\alpha)=I_{f^\JS_\alpha}(p:q)$\\
& for $f^\JS_\alpha(u) \eqdef (1-\alpha)f(\alpha u+1-\alpha)+\alpha	 f\left(\alpha+\frac{1-\alpha}{u}\right)$\\ 
\\
\\
Parameter distance & \\ \hline\hline
Bregman divergence & $B_F(\theta:\theta') \eqdef F(\theta)-F(\theta')-\inner{\theta-\theta'}{\nabla F(\theta')}$ \\
skew Jeffreys-Bregman divergence & $S_F^\alpha=(1-\alpha)B_F(\theta:\theta')+\alpha B_F(\theta':\theta)$\\
skew Jensen divergence & $J_F^\alpha(\theta:\theta') \eqdef (F(\theta)F(\theta'))_\alpha-F((\theta\theta')_\alpha)$\\
Jensen-Bregman divergence &  $\JB_F(\theta;\theta') = \frac{1}{2} \left( B_F\left(\theta:\frac{\theta+\theta'}{2}\right) + 
 B_F\left(\theta':\frac{\theta+\theta'}{2}\right) \right)=J_F(\theta;\theta')$.
\\
\\
Generalized Jensen-Shannon divergences & \\ \hline\hline
skew $J$-symmetrization & $J_D^\alpha(p:q) \eqdef (1-\alpha) D\left(p:q\right) +  \alpha D\left(q:p\right)$\\
skew $\JS$-symmetrization & $\JS_D^\alpha(p:q) \eqdef  (1-\alpha)D\left(p:(1-\alpha)p+\alpha q\right) +  \alpha D\left(q:(1-\alpha)p+\alpha q\right)$\\
skew $M$-Jensen-Shannon divergence & $\JS^{M_\alpha}(p:q) \eqdef  (1-\alpha)\KL\left(p:(pq)_\alpha^M\right)+\alpha\KL\left(q:(pq)_\alpha^M\right)$\\
skew $M$-$\JS$-symmetrization & $\JS^{M_\alpha}_D(p:q) \eqdef  (1-\alpha)D\left(p:(pq)_\alpha^M\right)+\alpha D\left(q:(pq)_\alpha^M\right)$
\\
$N$-Jeffreys divergence & $J^{N_\beta}(p:q) \eqdef  N_\beta(\KL\left(p:q\right),\KL\left(q:p\right))$\\
$N$-J D divergence  & $J^{N_\beta}_D(p:q) = N_\beta(D(p:q),D(q:p))$\\
skew $(M,N)$-$D$ JS divergence & ${\JS}^{M_\alpha,N_\beta}_D(p:q) \eqdef  N_\beta\left(D\left(p:(pq)_\alpha^M\right),D\left(q:(pq)_\alpha^M\right)\right)$\\
\end{supertabular}

\section{Symbolic calculations in {\sc Maxima}}\label{sec:cauchymax}

The program below (written in {\sc Maxima}, software available at \url{http://maxima.sourceforge.net/}) calculates the normalizer $Z$ for the harmonic $H$-mixtures of Cauchy distributions (Eq.~\ref{eq:ZH}).

\begin{verbatim}
assume(gamma>0);
Cauchy(x,gamma) := gamma/(%pi*(x**2+gamma**2));
assume(alpha>0);
h(x,y,alpha) :=  (x*y)/((1-alpha)*y+alpha*x);
assume(gamma1>0);
assume(gamma2>0);

m(x,alpha) := ratsimp(h(Cauchy(x,gamma1),Cauchy(x,gamma2),alpha));

/* calculate Z */
integrate(m(x,alpha),x,-inf,inf);
\end{verbatim}

\end{document}